\documentclass[aps,twocolumn,showpacs,preprintnumbers,amsmath,amssymb,superscriptaddress,prx,floatfix]{revtex4-2}

\DeclareMathAlphabet\mathbfcal{OMS}{cmsy}{b}{n}

\usepackage{graphicx}
\usepackage[]{hyperref}                          
\usepackage{xcolor}
\usepackage[english, nameinlink]{cleveref}          
\usepackage{xparse}                                 
\usepackage{bm}                                     
\usepackage{amsmath}           
\usepackage{soul}
\usepackage{verbatim}
\usepackage{placeins}                              

\renewenvironment{align}{
    \begin{equation}
    \begin{aligned}
}{
    \end{aligned}
    \end{equation}
    \ignorespacesafterend
}

\crefname{figure}{Fig.\@}{Figs.\@}
\Crefname{figure}{Figure}{Figures}
\crefname{equation}{Eq.\@}{Eqs.\@}
\Crefname{equation}{Equation}{Equations}
\crefname{table}{Tbl.\@}{Tbls\@}
\Crefname{table}{Table}{Tables}
\crefname{section}{Sec.}{Secs.}
\Crefname{section}{Section}{Sections}
\crefname{appendix}{App.\@}{Apps.\@}
\Crefname{appendix}{Appendix}{Appendix}

\hypersetup{
    colorlinks,
    linkcolor={black},
    citecolor={blue},
    urlcolor={blue!80!black}
}

\DeclareMathOperator{\arctanh}{arctanh}
\newcommand{\lp}{\left(}
\newcommand{\rp}{\right)}
\newcommand{\lsb}{\left[}
\newcommand{\rsb}{\right]}

\renewcommand*{\d}{\, \mathrm{d}}      

\NewDocumentCommand\pdv{m+g}{%
  \IfNoValueTF{#2}
    {\frac{\partial}{\partial {#1}}}
    {\frac{\partial {#1}}{\partial {#2}}}%
}
\newcommand{\mel}[3]{\langle #1  | #2 | #3\rangle}
\newcommand{\braket}[2]{\langle #1  | #2 \rangle}
\newcommand{\abs}[1]{\left| #1 \right|}

\allowdisplaybreaks

\begin{document}

\title{Nonequilibrium Relaxation and Odd-Even Effect in Finite-Temperature Electron Gases}

\author{Eric Nilsson}
\email{nieric@chalmers.se}
\affiliation{Department of Physics, Chalmers University of Technology, 41296 Gothenburg, Sweden}

\author{Ulf Gran}
\email{ulf.gran@chalmers.se}
\affiliation{Department of Physics, Chalmers University of Technology, 41296 Gothenburg, Sweden}

\author{Johannes Hofmann}
\email{johannes.hofmann@physics.gu.se}
\affiliation{Department of Physics, Gothenburg University, 41296 Gothenburg, Sweden}
\affiliation{Nordita, Stockholm University and KTH Royal Institute of Technology, 10691 Stockholm, Sweden}

\date{\today}

\begin{abstract}
Pauli blocking in Fermi liquids imposes strong phase-space constraints on quasiparticle lifetimes, leading to a well-known quadratic-in-temperature decay rate of quasiparticle modes at low temperatures. In two-dimensional systems, however, even longer-lived modes are predicted  (dubbed ``odd-parity'' modes) that involve a collective deformation of the Fermi distribution. Here, we present an efficient method to evaluate the full spectrum of relaxational eigenmodes of a Fermi liquid within kinetic theory. We employ this method to study the experimentally relevant case of a Fermi liquid with screened Coulomb interactions and map out the decay rates of quasiparticle modes beyond the asymptotic low-temperature limit up to the Fermi temperature, thus covering the entire temperature range of typical experiments. We confirm the existence of anomalously long-lived odd-parity modes and provide a comprehensive classification and detailed analysis of the relaxation spectrum. In particular, we find that (i)~the odd-parity effect in the decay rates extends to temperatures as large as $T=0.15T_F$, (ii)~there is only a small number of long-lived odd-parity modes, with an infinite number of remaining modes that show standard Fermi-liquid scaling, and (iii)~the ratio between the odd- and even-parity lifetimes is tunable with the Coulomb interaction strength, in addition to temperature, which reflects a difference in the microscopic relaxation mechanism of the modes. Our findings provide a comprehensive description of the nonequilibrium relaxation behavior of two-dimensional electron gases and bridge a significant gap in our understanding of these systems.
\end{abstract}

\maketitle

\section{Introduction}

While transport in conventional metals is governed by momentum-nonconserving relaxation processes like impurity or phonon scattering, it has recently become possible to create exceptionally clean two-dimensional electron gases in which binary electron-electron scattering dominates. The rate of such binary collisions is expected to scale with the square of the  temperature at low temperatures, \mbox{$\gamma \sim T^2/\hbar T_F$}~\cite{pines_theory_2018,baym_landau_1991}, where $T_F$ is the Fermi temperature, which points to the suppression of quasiparticle interactions at low temperatures and justifies the surprising effectiveness of a quasiparticle description of electron transport~\cite{sondheimer50}. It also implies that as the temperature is increased to a sizable fraction of the Fermi temperature~$T_F$, electron interactions become sufficiently strong in clean materials that the system crosses over from a ballistic or diffusive regime to an interaction-dominated hydrodynamic regime. Signatures of hydrodynamic electron behavior have now been reported in many  materials~\cite{bandurin_negative_2016,crossno_observation_2016,krishna_kumar_superballistic_2017,nam_electronhole_2017,berdyugin_measuring_2019,bandurin_fluidity_2018, braem_scanning_2018,gooth_thermal_2018,moll_evidence_2016,gusev_viscous_2018,ginzburg_superballistic_2021,aharonsteinberg22,kumar_imaging_2022,palm_observation_2024}, 
at temperatures that are typically $100$--$150$K for monolayer~\cite{bandurin_negative_2016} and bilayer~\cite{bandurin_fluidity_2018} graphene and $30$K for Ga[Al]As heterostructures~\cite{ginzburg_superballistic_2021}. Seen in relation to the Fermi temperature $T_F$ in these materials, more recent experiments even report hydrodynamic transport at temperatures as low as $T=0.02 T_F$~\cite{zeng24,estradaalvarez25,sarypov25}. 

\begin{figure}[t]
    \centering
    \includegraphics{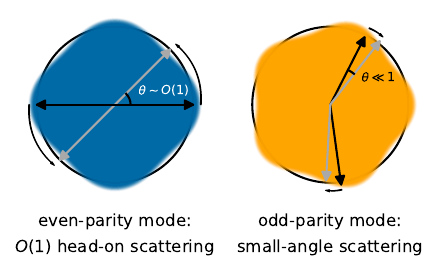}
    \caption{
    Microscopic origin of the odd-even effect in the relaxation of the quasiparticle distribution. Sketched are two nonequilibrium distributions in momentum space with a deformed Fermi surface and a small broadening of the Fermi edge of width~${\it O}(T)$, where the first distribution (blue) has even and the second (orange) odd parity. At low temperatures, relaxation corresponds to angular dynamics on the Fermi surface (black circle). Even-parity deformations of the quasiparticle distribution (blue) relax by head-on scattering, which gives the canonical Fermi-liquid dependence \mbox{$\gamma_{\text{even}} \sim (T/T_F)^2$}. Odd-parity deformations (orange), by contrast, relax by small-angle scattering, which leads to a much smaller relaxation rate at low temperatures \mbox{$\gamma_{\text{odd}} \sim (T/T_F)^4$}.}
    \label{fig:1}
\end{figure}

Fundamentally, the quadratic-in-temperature Fermi-liquid scaling of relaxation  rates at low temperatures is a consequence of a phase-space constraint that allows  interactions only between excitations close to the Fermi surface. Here, at low temperatures, the relaxation to the equilibrium distribution is described by angular dynamics on the Fermi surface. This is illustrated in \cref{fig:1} for a generic quasiparticle distribution $f(t, {\bf p})$ in momentum space (indicated by the shaded area) that deviates from the Fermi-Dirac distribution (indicated by the black circle that marks the Fermi energy). Energy and momentum conservation, together with the Fermi surface constraint in two dimensions, impose strong restrictions on possible relaxation channels, since binary forward or exchange scattering processes do not relax the distribution at this order. However, direct head-on processes, which create final states rotated by any angle on the Fermi surface, affect opposite points on the Fermi surface equally and can thus relax deformations that are parity even, i.e., that are described by the symmetric part of the distribution \mbox{$f_{\rm even} = [f({\bf p}) + f(-{\bf p})]/2$}. This type of process, exemplified for the blue Fermi surface deformation in~\cref{fig:1}, gives rise to the quadratic-in-temperature Fermi-liquid scaling.

Such a quadratic-in-temperature scaling of the relaxation rates of the {\it collective} Fermi surface deformation~$f_{\rm even}$ already represents a striking deviation from the expected relaxation rate of a single quasiparticle in two dimensions, which at low temperatures scales as \mbox{$\gamma_{\rm qp} \sim T^2 \ln (T_F/T)/(\hbar T_F)$}~\cite{hodges_effect_1971,giuliani_lifetime_1982} with an additional logarithmic temperature factor. Crucially, it is the relaxation rate of collective quasiparticle deformations (and not that of single quasiparticles) that will determine the magnitude of transport coefficients: Such transport coefficients describe the relaxation of a particular deformation of the quasiparticle distribution in response to an external perturbation, and will involve very different moments of the distribution function, with potentially very different relaxation rates. For example, the structure factor and bulk viscosity will depend on the $s$-wave density projection, $\int d\theta f({\bf p})$ (where $\theta$ is the angle of the ${\bf p}$ vector in polar coordinates), charge and heat currents on the $p$-wave current projection, $\int d\theta (\cos \theta, \sin \theta) \, f({\bf p})$, or the shear viscosity on the $d$-wave projection, $\int d\theta \sin 2 \theta \, f({\bf p})$, all with an additional specific dependence on the magnitude $p$ of the  momentum. It is therefore essential to gain a complete understanding of the full spectrum of relaxational modes in a Fermi liquid without making approximate relaxation-time assumptions or relying, for example, on self-energy calculations.

Odd-parity deformations, by contrast, which are described by the antisymmetric part of the distribution \mbox{$f_{\rm odd} = [f({\bf p}) - f(-{\bf p})]/2$}, are not expected to decay by head-on collisions but have a much longer lifetime~\cite{laikhtman_electron-electron_1992,gurzhi95,nilsson05}. On a microscopic level, such modes are predicted to relax in a different way compared to the even-parity decay by repeated simultaneous head-on and small-angle scattering~\cite{ledwith_angular_2019}, as is shown for the orange Fermi surface deformation in~\cref{fig:1}. This process is interpreted as subdiffusive electron dynamics on the Fermi surface, and it leads to a much smaller decay rate that scales with the fourth power of temperature at low temperatures,~\mbox{$\gamma_{\text{odd}} \sim T^4/\hbar T_F^3$}~\cite{ledwith_hierarchy_2019,hofmann_anomalously_2023}. The argument relies on the presence of a sharp Fermi surface, and the odd-even effect in the quasiparticle relaxation is thus expected to vanish at higher temperatures where the Fermi distribution broadens. The intricate dependence of decay rates on the odd or even parity of deformations shows that there are many aspects of interaction-dominated Fermi liquids that remain to be understood.

In particular, the odd-even effect in the quasiparticle lifetimes implies that interacting two-dimensional Fermi liquids are expected to show a much richer behavior with temperature than a simple crossover between a ballistic and a hydrodynamic regime. To see this, recall that in the hydrodynamic regime interactions are so strong that typical lifetimes are much shorter compared to the timescales at which the system is observed, such that only long-lived modes (i.e., modes corresponding to densities of conserved quantities) contribute. However, the vastly different odd and even decay rates allow for an intermediate regime at timescales \mbox{$\gamma^{-1}_{\text{even}} \ll t \ll \gamma^{-1}_{\text{odd}}$}, where even-parity modes have decayed but odd-parity modes must be included in an extended hydrodynamic picture, sometimes referred to as tomographic transport. Such a new transport regime could be accessed in experiments at intermediate temperature scales, and there is currently intense research to establish signatures of this odd-even effect in transport~\cite{hofmann_anomalously_2023,ledwith_hierarchy_2019,ledwith_angular_2019,ledwith_tomographic_2019,hofmann_collective_2022,kryhin_collinear_2023,hofmann_nonlinear_2023}. Excitingly, recent experimental works report signatures of tomographic flow and long-lived odd-parity modes~\cite{zeng24,levitov24,moisenko25}.

However, even when discussing the basic relaxation times of a Fermi liquid, many questions are still open. First, since the interaction-dominated Fermi liquid regime exists only once the temperature reaches a sizable fraction of the Fermi temperature (where interactions dominate over impurity scattering), the temperature dependence of the decay rates must be understood beyond the asymptotic low-temperature scaling discussed above. The central question is whether the odd-even effect will persist in an experimentally relevant regime. In a previous study we have computed the Fermi-liquid decay rates with a constant interaction potential, for which the odd-even effect exists below temperatures \mbox{$T\lesssim 0.15T_F$}~\cite{hofmann_anomalously_2023}. However, quantitative predictions for the experimentally relevant case of a screened Coulomb potential, for example, are missing. Such an interaction takes the form
\begin{equation}
    V(\mathbf{k}) = 
    \frac{2 \pi e^2}{k + k_{\rm TF}} ,
    \label{eq:Coulomb_matel}
\end{equation}
where \mbox{$k_{\rm TF} = 2m^* e^2/\hbar^2$} is the Thomas-Fermi wave vector, with a dimensionless interaction strength \mbox{$r_s = k_{\rm TF}/(2\sqrt{\pi n})$}, where $n$ is the electron density and $m^*$ the effective mass~\cite{giuliani_quantum_2005}. A second open point is that, for any interaction, the odd-even effect is discussed so far in terms of lowest-lying modes, and the full excitation spectrum of a Fermi liquid within kinetic theory has not been established. A central question here is if there exists an isolated set of modes with anomalously long lifetimes, or if odd-parity modes are generically long-lived. To identify the long-lived modes would be particularly relevant for effective beyond-hydrodynamic theories. Third, the Coulomb interaction,~\cref{eq:Coulomb_matel}, provides an entirely separate modification of small-angle scattering processes due to the increased matrix element at small momentum transfer $k$. Such an enhancement typically leads to logarithmic corrections to the Fermi-liquid temperature scaling, but is expected to affect odd and even modes in a different way, as the decay of odd-parity modes is already controlled by small-angle processes. The magnitude of  the odd-even effect could thus be tuned by the interaction strength.

In this paper, we answer these questions and provide a comprehensive discussion of collective quasiparticle lifetimes in a two-dimensional electron gas including the screened Coulomb interaction~\cref{eq:Coulomb_matel}. We observe the odd-parity effect in quasiparticle lifetimes over a large and experimentally detectable temperature range below~$T_F$. Remarkably, we find that the odd parity of the mode is not sufficient to have an anomalously suppressed decay rate: We classify every long-lived odd-parity mode by a corresponding symmetry of the Fermi surface deformation and find that for each class the modes separate by a large gap from higher-order modes in the same sector, which are found to decay with standard Fermi-liquid scaling. Furthermore, we find that odd-parity modes do not follow the standard Fermi liquid ${\it O}(r_s^2)$ dependence on the interaction strength, which reflects the different microscopic relaxation dynamics and which indeed shows that the separation between modes of even and odd parity is tunable with the interaction strength.

For definiteness, we focus on simple Fermi liquids with a parabolic single-particle dispersion \mbox{$\varepsilon({\bf p}) = \hbar^2p^2/2m^*$}, as is appropriate, for example, for GaAs or doped graphene. In general, the odd-even effect exists as long as the Fermi surface is symmetric under point inversion \mbox{${\bf p} \to - {\bf p}$}, which implies \mbox{$\varepsilon({\bf p}) = \varepsilon(-{\bf p})$}, which is the case for any time-reversal invariant material. The key quantities that we compute are the relaxation rates of collective deformations in the quasiparticle distribution. These relaxation rates describe how small deviations from the equilibrium Fermi-Dirac distribution \mbox{$f_{0}(\mathbf{p}) =  1/(\exp{[\beta(\varepsilon(\mathbf{p}) - \mu)] + 1)}$} (where \mbox{$\beta = 1/T$} with \mbox{$k_B = 1$} is the inverse temperature and $\mu$ the chemical potential) relax back to equilibrium due to binary collisions with the interaction potential~\cref{eq:Coulomb_matel} (cf.~\cref{fig:1}). Formally, each collective decay mode is an eigenfunction of the collision integral,
\begin{align}
    &\hat{\mathcal{J}}[f(t,\mathbf{p}_1)]  = - 
    \int \frac{\d \mathbf{p}_2 \d \mathbf{p}'_1 \d \mathbf{p}'_2}{(2\pi)^6} \, 
    W(\mathbf{p}_1', \mathbf{p}_2' | \mathbf{p}_1 \mathbf{p}_2) \\ &\quad \times  \Bigl[ f(\mathbf{p}_1) f(\mathbf{p}_2) [1 - f(\mathbf{p}_1')] [1-f(\mathbf{p}_2')] \\
    &\quad \qquad - f(\mathbf{p}_1') f(\mathbf{p}_2') [1- f(\mathbf{p}_1)] [1-f(\mathbf{p}_2)] \Bigr] ,
    \label{eq:collision_def}
\end{align}
where $W(\mathbf{p}_1', \mathbf{p}_2' | \mathbf{p}_1 \mathbf{p}_2)$ is the Coulomb scattering matrix that enforces energy and momentum conservation. Here, the first term in the square brackets accounts for a loss of quasiparticles with wave vector $\mathbf{p}_1$ due to scattering off another quasiparticle with wave vector $\mathbf{p}_2$ as \mbox{$\mathbf{p}_1 + \mathbf{p}_2 \to \mathbf{p}_1' + \mathbf{p}_2'$}, while the second term describes the reverse gain as two quasiparticles scatter into the states with momenta $\mathbf{p}_1$ and $\mathbf{p}_2$. The collision integral vanishes identically if $f(t, \mathbf{r}, \mathbf{p})$ is equal to the Fermi-Dirac distribution, and we are interested in small deviations \mbox{$f(t, \mathbf{p}) = f_{0}(\mathbf{p}) + \delta f(t, \mathbf{p})$}, 
which we parametrize as
\begin{align}
    \delta f(t, \mathbf{p}) &= \lp - T \frac{\partial f_{0}}{\partial \varepsilon}\rp \psi(t, \mathbf{p}) \\
    &= f_{0}(\mathbf{p}) [1- f_{0}(\mathbf{p})] \psi(t, \mathbf{p}).
    \label{eq:f_pert_def}
\end{align}
The function $\psi(t,{\bf p})$ that parametrizes the quasiparticle deviation can be thought of as a time- and momentum-dependent variation of the chemical potential. The symmetry of $\psi$ in momentum space then sets the symmetry of the Fermi surface deformation. For the particular case of a circular Fermi surface, the perturbations may be expanded in angular harmonics labeled by an angular mode number $m$ as
\begin{equation}
    \psi_m (p) = \int_0^{2 \pi} \frac{\d \theta}{2 \pi} e^{-i m \theta} \psi (\mathbf{p}),
    \label{eq:psi_m}
\end{equation}
with $\theta$ the polar angle in the $\mathbf{p}$ plane and $p$ the magnitude of the momentum. 
\begin{figure}[t]
    \centering
    \includegraphics{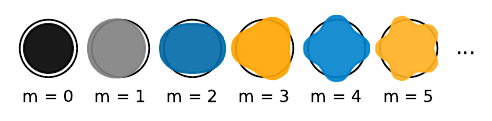}
    \caption{Symmetry of Fermi surface deformations ordered by the angular harmonic index $m$ [cf. \cref{eq:psi_m}]. The black circle indicates the Fermi surface. Note that the full relaxational eigenmodes of the collision integral, \cref{eq:collision_def}, are labeled by the angular symmetry,  but they are in general not sharp deformations of the Fermi distribution as sketched here [i.e., \cref{eq:f_pert_def} with \mbox{$\psi = {\rm const}$}]. Instead, they have a thermal and interaction-induced broadening, where higher eigenmodes have an increasing number of radial nodes in the function $\psi_m(p)$ [cf. Eqs.~\eqref{eq:basis_expansion} and~\eqref{eq:lowT_exact_polys} for the low-temperature case]. 
   Deformations with \mbox{$m=0$} symmetry determine the density response and bulk viscosity, \mbox{$m=1$} to charge and heat currents, and \mbox{$m=2$} to the shear viscosity, for example, and higher modes contribute to a finite wave vector response.
    }
    \label{fig:fermi_m_illustration}
\end{figure}
A constant $\psi_m$ then describes a rigid deformation of the Fermi surface, and an additional energy dependence accounts for a thermal or interaction-induced broadening. The symmetry of different Fermi surface deformations is illustrated in~\cref{fig:fermi_m_illustration} for a rigid deformation of the zero-temperature Fermi surface. For rotationally invariant interactions, where decay rates $\gamma_m$ in different angular channels decouple, odd-parity modes are described by odd $m$ (orange color in \cref{fig:fermi_m_illustration}) and even-parity modes by even $m$ (blue color in \cref{fig:fermi_m_illustration}). The different deformations that determine different transport coefficients discussed above correspond to different angular-mode sectors; for example, \mbox{$m=0$} for the structure factor and bulk viscosity, \mbox{$m=1$} for charge and heat currents, and \mbox{$m=2$} for the shear viscosity~\cite{gran_shear_2023-1}. Furthermore, a coupling to higher angular modes exists in the response to a perturbation at finite wavelength. Likewise, collective sound or plasmon modes contain a superposition of different angular components~\cite{hofmann_collective_2022}. Beyond the angular mode decomposition, the functional dependence of an eigenmode $\psi_m(p)$ on the momentum (corresponding to an energy dependence of the mode) does not separate further, and in general there is an infinite set of eigenmodes in each angular momentum channel.

The key advance that allows us to present results beyond the asymptotic low-temperature limit described in the classical literature~\cite{abrikosov_theory_1959} is a general basis expansion for the energy dependence of eigenmodes, which allows us to compute the entire spectrum of decay rates across a wide temperature range for modes of different angular harmonics. We note that our framework extends classical works on the basis expansion of classical gases by Enskog~\cite{enskog_kinetische_1917}, Chapman and Cowling~\cite{chapman_mathematical_1990}, and others~\cite{grad_kinetic_1949,burnett_distribution_1935}, dating back to the early 1900s, and the formalism outlined in this work can be used to describe interacting Fermi liquids beyond the asymptotic low- or classical high-temperature limit. We focus here on temperatures \mbox{$T\leq T_F$} relevant for the odd-even effect, and we leave the exploration of more general potentials $W$ in \cref{eq:collision_def} (such as the full RPA-Lindhard function, or adapted to capture the Wigner crystal transition) to future work.

\begin{figure*}[t]
    \centering
    \scalebox{1.03}{\includegraphics{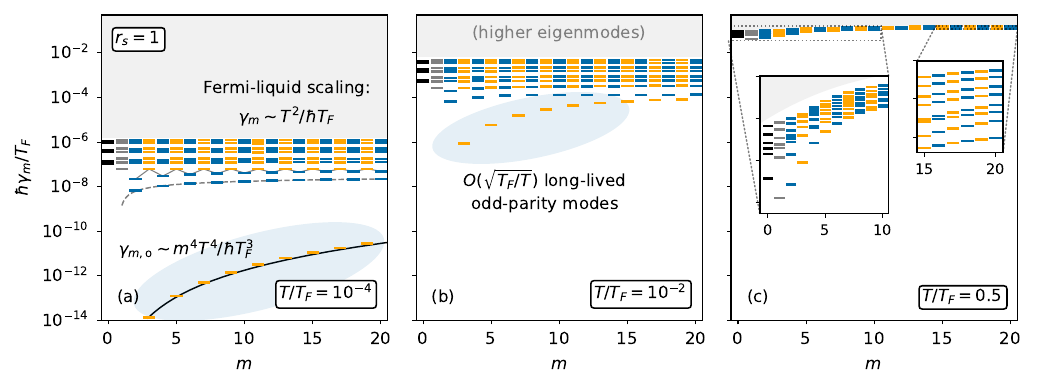}}
    \caption{Spectra of decay rates $\gamma_m$ as a function of  angular mode number $m$ for \mbox{$r_s = 1$} at three different temperatures (a)~\mbox{$T/T_F=10^{-4}$}, (b)~\mbox{$T/T_F = 10^{-2}$}, and (c)~\mbox{$T/T_F=0.5$}. We show the lowest eight relaxational eigenmodes for each angular channel, where compressional modes (\mbox{$m=0$}) are black, current modes (\mbox{$m=1$}) gray, and remaining even-parity modes blue and odd-parity modes orange, respectively, and we indicate higher modes by the gray shaded area. We use the same range in each plot to compare spectra at different temperatures. As the temperature decreases, Pauli blocking becomes increasingly important with a phase-space constraint on scattering that increases the lifetime, where the magnitude of most relaxation rates asymptotically follows a Fermi-liquid scaling with \mbox{$\gamma_m \sim T^2/(\hbar T_F)$}. Crucially, however, there is an odd-even effect in the relaxation rates at low temperatures, with a band of isolated odd-parity modes (the number of which increases at low temperature as \mbox{$\sim \sqrt{T_F/T}$}) that decouple from the remaining spectrum with anomalously long lifetimes and an asymptotic scaling \mbox{$\gamma_m \sim m^4 T^4/(\hbar T_F^3)$} [highlighted by the blue shaded area in (a) and (b)].
    }
    \label{fig:low_T_leveldiag}
\end{figure*}

This paper is organized as follows. We begin in \cref{sec:results} by presenting and collecting the central results for the collective quasiparticle decay rates. Section~\ref{sec:oddeven} establishes the odd-even effect and discusses the dependence on the angular index $m$. In particular, we show that there exists an isolated band of long-lived odd-parity modes that is separated by a gap in the same angular sector from the remaining odd-parity modes with Fermi-liquid behavior. This is followed in~\cref{sec:temperaturedependence} by a detailed study of the temperature dependence of the relaxational eigenmodes. Section~\ref{sec:interaction} studies the dependence of the decay rates on the interaction strength and establishes that the odd-parity modes are nearly independent of it. \Cref{sec:background} outlines in detail the numerical diagonalization of the linearized collision integral (\cref{sec:linearized_collision}), which is based on a basis expansion of  quasiparticle deformations (\cref{sec:basis_expansion}). \Cref{sec:convergence} provides details on an efficient numerical implementation followed by a discussion of zero modes in \cref{sec:zeromodes}. The paper concludes with a summary and outlook in \cref{sec:conclusion}. Details of analytical calculations are collected in four appendixes.

\section{Results}\label{sec:results}

In this section, we present results for the relaxation spectrum of quasiparticle deformations as obtained from our numerical diagonalization of the collision integral, \cref{eq:collision_def}, which is the main result of this paper. \Cref{sec:oddeven} compares the spectrum of relaxation rates for different temperatures  and demonstrates that there is an odd-even effect in the lowest eigenmodes but not for higher excitations. This is followed in~\cref{sec:temperaturedependence} by a discussion of the temperature dependence of the lowest eigenmodes, establishing both Fermi-liquid as well as anomalous temperature scaling. Finally, in~\cref{sec:interaction}, we show that the lowest even- and odd-parity modes also differ in their functional dependence on the interaction strength, which reflects the different microscopic relaxation mechanisms of the lowest odd- and even-parity modes and points to a tunability of the odd-even effect with interactions. The general solution method for the Fermi-liquid collision integral with binary scattering used to obtain these results is discussed in detail in \cref{sec:background}, and analytical limits are discussed in Appendixes~\ref{app:lowTdetails} and~\ref{app:low_rs_limit}. 

\subsection{Relaxation spectrum and the odd-even effect}\label{sec:oddeven}

We begin by discussing the general structure of the relaxation spectrum. \Cref{fig:low_T_leveldiag} shows level diagrams of the relaxation rates $\gamma_m$ for \mbox{$r_s=1$} ordered by the angular index \mbox{$0\leq m \leq 20$} at three different temperatures \mbox{$T/T_F=10^{-4}, 10^{-2}$}, and $0.5$ [Figs.~\ref{fig:low_T_leveldiag}(a) to~\ref{fig:low_T_leveldiag}(c)]. In each angular harmonic channel, we show the lowest eight eigenvalues and use the same axes range to compare spectra at different temperatures. Higher relaxational eigenmodes are indicated by the shaded gray area. Relaxation rates for negative harmonics $-m$ are equal to positive $m$ and not shown here. We mark even-parity modes in blue (even \mbox{$m\geq 2$}) and odd-parity modes in orange (odd \mbox{$m\geq 3$}). The \mbox{$m=0$} and \mbox{$m=1$} sectors are black and gray, respectively, since they contain zero modes for which the discussion of the odd-even effect does not apply. (The \mbox{$m=0$} sector contains two zero modes, arising from particle number and energy conservation, while the \mbox{$m=\pm 1$} sectors each contain one zero mode, arising from current conservation, cf.~\cref{sec:zeromodes}.) These modes are therefore not visible in~\cref{fig:low_T_leveldiag} since their decay rates vanish, while relaxation rates of higher modes in the same sector are finite. All temperatures shown are in the degenerate low-temperature regime below $T_F$, with a Fermi surface that increasingly softens when going from the left-hand to the right-hand panel and a concomitant weakening of Pauli blocking.

The first striking feature of~\cref{fig:low_T_leveldiag} is the separated lowest band of orange odd-parity modes, which have anomalously small relaxation rates (highlighted by the blue shaded area). These are the long-lived odd-parity modes discussed in the Introduction. As is apparent from the figure, these modes have a strong $m$ dependence and are seen to join the remaining spectrum at large~$m$. The number of decoupled modes increases as the temperature is lowered, roughly as \mbox{$\sqrt{T_F/T}$}, with an increasing suppression of the relaxation rate compared to higher-lying modes. The second remarkable feature is that the odd-even effect exists only for the lowest relaxational eigenmodes: The overall magnitude of all higher eigenmodes decreases at smaller temperatures with a temperature dependence that in scale follows a standard Fermi-liquid form \mbox{$\gamma_m \sim T^2/\hbar T_F$} in all angular sectors, where the overall level spacing also has Fermi-liquid scaling and increases at low temperatures. These higher modes show no  pronounced anomalous behavior, with the only exception visible for the second-lowest relaxation rates at the lowest temperature \mbox{$T/T_F=10^{-4}$}, which show a weak remaining odd-even staggering [thin gray line in~\cref{fig:low_T_leveldiag}(a)]. In addition, higher-lying modes are almost independent of the angular index, but they do show a pair bunching for relaxation rates in the same angular sector. At the highest temperature \mbox{$T/T_F=0.5$} shown [\cref{fig:low_T_leveldiag}(c)], the effect of Pauli blocking is reduced and the gas crosses over to a nondegenerate regime. As is seen most clearly for the highest values of $m$ in~\cref{fig:low_T_leveldiag}(c) (right-hand inset), the level separation between the modes is then increasingly equidistant. At this temperature, there is no longer an odd-even effect, with Pauli blocking affecting modes with smaller angular index $m$ more strongly compared to large-$m$ modes (left-hand inset). 

\begin{figure*}[t]
    \centering
    \scalebox{1.03}{\includegraphics{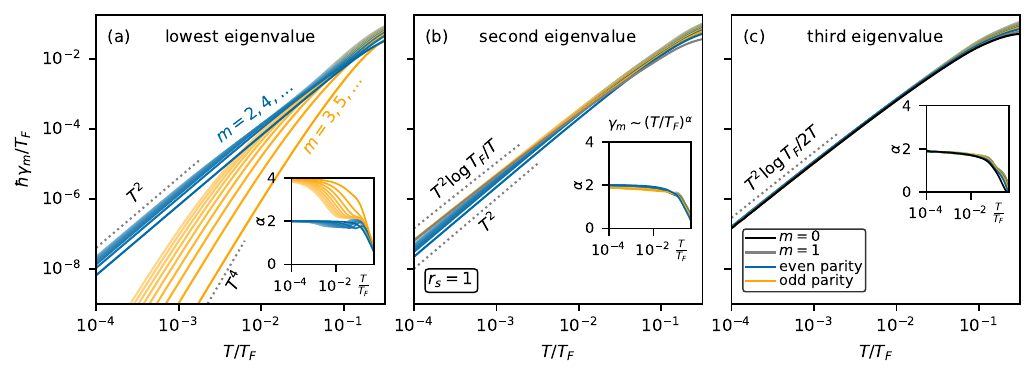}}
    \caption{Relaxation rates as a function of temperature for the (a) lowest, (b) second-lowest, and (c) third-lowest relaxational eigenmodes (left to right) in each angular sector up to \mbox{$m \leq 20$}. We show compressional modes [\mbox{$m=0$}, (c)] in black, current modes [\mbox{$m=1$}, (b) and (c)] in gray, and remaining even-parity modes in blue and odd-parity modes in orange, respectively. The insets show the local temperature scaling exponent $\alpha$ for each mode, \mbox{$\gamma_m \sim (T/T_F)^\alpha$}, as determined from a logarithmic derivative of the numerical data in the main figures. The relaxation rate of all modes decreases with decreasing temperature, reflecting the enhanced lifetime due to Pauli blocking. For blue even-parity modes in (a) and all  higher eigenmodes [(b) and (c)], the relaxation rates are suppressed at low temperatures with standard Fermi-liquid scaling, \mbox{$\gamma_m \sim T^2/(\hbar T_F)$}, corresponding to \mbox{$\alpha=2$}. The odd-even effect is present only in the lowest eigenmode of each angular momentum sector (a): Here, with decreasing temperature, a successively   increasing number of order \mbox{${\it O}(\sqrt{T_F/T})$} of odd-parity modes decouple from the Fermi-liquid scaling with an asymptotic temperature dependence  \mbox{$\gamma_m \sim T^4$}, corresponding to \mbox{$\alpha=4$}.}
    \label{fig:allrates_temperature}
\end{figure*}

Quantitatively, we find that the anomalous lowest odd-parity modes at low temperatures are very well described by
\begin{equation}
    \gamma_{m, \text{odd}} = \frac{4\pi^3 T_F}{15 \hbar} \abs{\bar{V}}^2 \lp\frac{T}{T_F}\rp^4 m^4 ,
    \label{eq:lowT_odd_gamma}
\end{equation}
shown as a black line in \cref{fig:low_T_leveldiag}(a). The expression applies for angular indices \mbox{$m \lesssim \sqrt{T_F/T}$}, where as discussed for higher angular momenta the odd-even effect disappears and odd-parity modes show standard Fermi-liquid scaling. 
In~\cref{eq:lowT_odd_gamma}, $\abs{\bar{V}}^2 = (m^*/2 \pi \hbar^2)^2 [V^2(0) + V^2(2k_F) - V(0)V(2k_F)]$ is the (dimensionless) matrix element of~\cref{eq:Coulomb_matel} symmetrized over the direct and exchange scattering channel. 
[Formally, \cref{eq:lowT_odd_gamma} is asymptotically valid, and the plot in \cref{fig:low_T_leveldiag}(a) uses the local temperature scaling exponent \mbox{$(T/T_F)^{3.965}$}, which applies for \mbox{$T/T_F = 10^{-4}$}, and evaluates the Coulomb matrix element at a small momentum transfer \mbox{$k = 0.2 k_F$}; see the next section for details.] The quartic \mbox{$m^4$} scaling of \cref{eq:lowT_odd_gamma}, which is very accurately seen in the data at low temperature [cf.~\cref{fig:low_T_leveldiag}(a)], reflects subdiffusive behavior~\cite{ledwith_angular_2019}: Written as an effective model for the angular relaxation of the odd-parity distribution \mbox{$f_{\rm odd} = \sum_{m \, {\rm odd}} f_m e^{im\theta}$}, \cref{eq:lowT_odd_gamma} describes a dynamics,
\begin{align}
    \partial_t f_{\rm odd} &= - \gamma_{\rm odd} \, \partial_\theta^4 f_{\rm odd} ,
\end{align}
with $\gamma_{\rm odd} = \gamma_{m\text{, odd}}|_{m=1}$ the $m$ coefficient of \cref{eq:lowT_odd_gamma}. The equation implies that an initial localized deformation of the quasiparticle distribution spreads as \mbox{$t^{1/4}$}, which is much slower than diffusion (which would have \mbox{$\gamma_m \sim m^2$} with an exponent~$1/2$), hence subdiffusive. \Cref{eq:lowT_odd_gamma} is in agreement with an analytical prediction by~\textcite{ledwith_hierarchy_2019}. Reference~\cite{ledwith_hierarchy_2019} contains an additional logarithmic dependence on $m$ that is not present in our results.

Turning to the lowest even-parity modes, they follow a standard quadratic-in-temperature Fermi-liquid scaling 
\begin{equation}
    \gamma_{m \text{, even}} \approx \frac{2 \pi}{3\hbar} \frac{T^2}{T_F} r_s^2 \biggl[ \log\biggl( 1 + \frac{\sqrt{2}}{r_s}\biggr) - \frac{\sqrt{2}}{\sqrt{2}+r_s}\biggr] \frac{\ln m \phi}{\ln 2 \phi} ,
    \label{eq:lowT_even_pred}
\end{equation}
shown as a gray dashed line in~\cref{fig:low_T_leveldiag}(a). This result is derived in~\cref{sec:interaction} and Appendix~\ref{app:lowTdetails}, and is applicable for \mbox{$r_s \gtrsim 0.5$} [cf.~\cref{sec:interaction}].
The result is valid to leading logarithmic order, and we include a parameter $\phi$ to parametrize the subleading corrections; for \mbox{$r_s = 1$} and \mbox{$T/T_F = 10^{-4}$} as applies to~\cref{fig:low_T_leveldiag}(a), we have \mbox{$\phi=1.23$}. Note that~\cref{eq:lowT_even_pred} has a logarithmic correction in $m$ but not in temperature, in contrast to a quasiparticle relaxation time obtained from the imaginary part of the self-energy~\cite{giuliani_lifetime_1982,zheng_coulomb_1996,sarma_know_2021}. Formally, such a single-quasiparticle excitation corresponds to a localized excitation at a fixed momentum above the Fermi surface, and is a superposition of modes with arbitrarily high angular index. 
\subsection{Temperature dependence of relaxation rates}
\label{sec:temperaturedependence}
Having established the odd-even effect and the scaling of relaxation rates with the angular harmonic index $m$, we proceed in this section to discuss the temperature dependence of the eigenmodes. \Cref{fig:allrates_temperature} shows the lowest [\cref{fig:allrates_temperature}(a)], second-lowest [\cref{fig:allrates_temperature}(b)], and third-lowest [\cref{fig:allrates_temperature}(c)] eigenvalues (left- to right-hand panel) at a fixed interaction strength \mbox{$r_s=1$} as a function of temperature. Different lines indicate different angular indices, and we again use a color coding that shows odd-parity modes in orange and even-parity modes in blue. In addition, the first nonzero mode in the \mbox{$m=0$} sector (which has two zero modes from particle and energy conservation) is included in black in \cref{fig:allrates_temperature}(c), and the first two nonzero modes in the \mbox{$m=1$} sector (which has one zero mode from momentum conservation) are shown in gray in Figs.~\ref{fig:allrates_temperature}(b) and~\ref{fig:allrates_temperature}(c).

As established in the previous subsection, the lowest modes [\cref{fig:allrates_temperature}(a)] show the odd-even effect: In the degenerate regime below the Fermi temperature, \mbox{$T\leq T_F$}, where the Fermi statistics becomes important and Pauli blocking restricts the phase space for quasiparticle scattering, the blue even-parity modes follow the quadratic-in-temperature Fermi-liquid scaling, asymptotically described by \cref{eq:lowT_even_pred} with a weak logarithmic-in-$m$ dependence of the magnitude. By contrast, the decay rates of odd-parity modes are much more strongly suppressed, and asymptotically described by~\cref{eq:lowT_odd_gamma} with a pronounced $m^4$ scaling with the angular index. Note that the successive decoupling of an increasing number \mbox{$m \lesssim \sqrt{T_F/T}$} of anomalous odd-parity modes [also visible in Figs.~\ref{fig:low_T_leveldiag}(a) and~\ref{fig:low_T_leveldiag}(b)] is directly apparent in \cref{fig:allrates_temperature}(a): At higher temperatures, the orange odd-parity modes follow the blue Fermi-liquid modes before they cross over into the asymptotic odd-parity scaling described by~\cref{eq:lowT_odd_gamma} at low temperatures. The temperature at which the odd-parity modes deviate from Fermi-liquid scaling is strongly suppressed with increasing mode index $m$. To illustrate this further, we show in the inset of \cref{fig:allrates_temperature}(a) the exponent of the logarithmic derivative of $\gamma_m$ with respect to temperature, which gives a local scaling exponent \mbox{$\gamma_m \sim (T/T_F)^\alpha$}: Even-parity modes follow Fermi-liquid scaling with \mbox{$\alpha=2$}, while odd-parity modes reach the \mbox{$\alpha=4$} scaling only at asymptotically low temperatures, with higher angular indices even having an intermediate plateau at \mbox{$\alpha\approx 2$}. The large separation between odd and even modes is seen to remain up to \mbox{$T \lesssim 0.15 T_F$}.

The absence of a pronounced odd-even staggering in the decay rates for the higher eigenmodes noted in the previous section is clearly visible in the plots for the second-lowest [\cref{fig:allrates_temperature}(b)] and the third-lowest [\cref{fig:allrates_temperature}(c)] eigenvalues. Leaving aside a small logarithmic spread, the eigenmodes are almost independent of the angular index and follow Fermi-liquid scaling at low temperatures. The higher eigenvalues in the \mbox{$m=0$} and \mbox{$m=1$} sectors, which are not zero modes, also follow the same Fermi-liquid scaling as the rest of the higher even and odd modes. Interestingly, when considering the local scaling exponent shown in the insets, we find an improved fit with an additional logarithmic correction \mbox{$\gamma_m \sim (T/T_F)^2 \log T_F/ T$} and \mbox{$\gamma_m \sim (T/T_F)^2 \log T_F/ 2T$} in the second- and third-lowest sectors, respectively, a result which is reminiscent of a self-energy calculation~\cite{giuliani_lifetime_1982}.
\subsection{Interaction-induced enhancement of the odd-even effect}\label{sec:interaction}
Contrasting the low-temperature forms~\cref{eq:lowT_odd_gamma,eq:lowT_even_pred} for the decay rates of odd- and even-parity modes, it is apparent that even-parity modes have a complicated dependence on the interaction strength $r_s$ while the odd-parity scaling depends on the matrix element~$|\bar{V}|^2$, which has a much weaker $r_s$ dependence. This dependence is linked to the fact that odd-parity modes relax by repeated small-angle scattering, for which the Coulomb interaction~\cref{eq:Coulomb_matel} is overscreened and weakly dependent on~$r_s$~\cite{giuliani_quantum_2005,lux13}, whereas even-parity modes involve momentum transfers of order \mbox{${\it O}(k_F)$}, which depends on the density and thus on $r_s$. In this section, we discuss this dependence on the interaction strength in more detail. We confirm that the interaction strength strongly affects the strength of even-parity damping but not that of odd-parity modes, such that the odd-even effect is widely tunable with interactions.

\begin{figure}[t]
    \centering
    \includegraphics{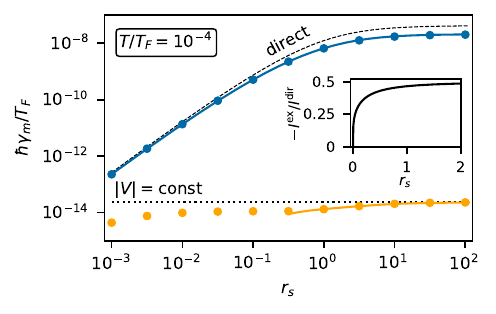}
    \caption{Lowest \mbox{$m = 2$} decay rate (blue points) and \mbox{$m=3$} decay rate (orange points) at \mbox{$T/T_F = 10^{-4}$} as a function of the Coulomb interaction strength $r_s$. The continuous blue line is the analytical even-parity result~\cref{eq:gamma_m2_exact}, and the dashed black line includes only the direct contribution in \cref{eq:direct_exchange_m2}. The latter expression becomes exact only at asymptotically small $r_s$ as illustrated in the inset, which shows the ratio between the exchange and direct scattering contributions. The continuous orange line is the exact odd-parity result \cref{eq:lowT_odd_gamma} with a momentum-dependent Coulomb scattering amplitude, and the dashed black line is the  decay rate for a constant matrix element at large~$r_s$ for comparison. Overall, there is a strong odd-even effect for all $r_s$, but the separation in the decay rates is strongly dependent on the interaction strength~$r_s$.
    }
    \label{fig:rs_dependence_lowT}
\end{figure}

\Cref{fig:rs_dependence_lowT} shows the decay rates of the lowest-lying \mbox{$m=2$} modes (blue points) and \mbox{$m = 3$} modes (orange points), respectively, as a function of the interaction strength~$r_s$ at a fixed small temperature~\mbox{$T/T_F = 10^{-4}$}. We focus here on the dependence of the decay rates on the Coulomb interaction parameter and neglect possible low-temperature phase transitions at large $r_s$ out of the Fermi-liquid phase~\cite{tanatar89,jamei05}. Indeed, while the even-parity decay rate varies by several orders of magnitude, the odd-parity rates remain almost constant. 
Discussing the odd-parity modes first, we show as a black dotted line the asymptotic limit of \cref{eq:lowT_odd_gamma} for large $r_s$, where the matrix element in \cref{eq:lowT_odd_gamma} is \mbox{$|\bar{V}|^2 = 1/4$}, which is in excellent agreement with the numerics. 
At decreasing $r_s$, a small decrease in the odd-parity relaxation rate is captured by the $r_s$ dependence of the matrix element and by allowing a small nonzero momentum transfer $k$ [\mbox{$k = 0.2 k_F$} for \mbox{$r_s=1$}], which indicates a small deviation from strict small-angle scattering processes that contribute to the relaxation. The fit of $\gamma_{m=3}$ from~\cref{eq:lowT_odd_gamma} is shown in~\cref{fig:rs_dependence_lowT} as a solid orange line, and is again in excellent agreement with the data in the range \mbox{$1\leq r_s \leq 100$}. Indeed, at the temperature $T/T_F = 10^{-4}$ shown in the figure, this is the same $r_s$ interval over which we find odd-parity temperature scaling with \mbox{$\alpha \approx 4$}. For smaller \mbox{$r_s \lesssim 1$} the scaling exponent $\alpha$ begins to decrease from its low-density value \mbox{$\alpha = 4$}.  The interaction strength thus not only controls the relative strength of the odd-even effect but also the onset into the asymptotic odd-parity $T^4$ regime. 

For the even-parity decay rate at low temperatures, we obtain the following analytical result valid for \mbox{$(T/T_F)\lesssim r_s$} (cf. Appendix~\ref{app:lowTdetails}):
\begin{equation}
    \gamma_{m\text{, even}} = \frac{4 \pi}{3 \hbar}\frac{T^2}{T_F} I_m, \quad I_m = I_{m}^{\text{dir}} + I_{m}^{\text{ex}},
    \label{eq:gamma_m2_exact}
\end{equation}
where, for \mbox{$m=2$},
\begin{align}
    I_{m=2}^{\text{dir}} &= r_s^2 \lsb \log \lp 1 +\frac{\sqrt{2}}{r_s}\rp-\frac{\sqrt{2}}{\sqrt{2}+r_s}\rsb , \\
    I_{m=2}^{\text{ex}} &=-\frac{\pi r_s^2}{4} + \frac{r_s^4}{2(r_s^2-1)} \Bigg[ \log\lp 1+ \frac{\sqrt{2}}{r_s}\rp \\
    & \qquad - \frac{\sqrt{2-r_s^2}}{r_s}\arctanh \lp \sqrt{1-r_s^2/2}\rp \Bigg] .
    \label{eq:direct_exchange_m2}
\end{align}
This result sets the low-temperature quadratic scaling of the shear viscosity~\cite{gran_shear_2023-1}. The two terms in \cref{eq:gamma_m2_exact} are the direct and exchange contributions, and we show the full expression as a continuous blue line in \cref{fig:rs_dependence_lowT}, in excellent agreement with the numerical results (blue points). For comparison, we show as a dashed black line only the direct contribution, which sets the full result only in the extreme high-density limit \mbox{$r_s \ll 1$}.  We show the ratio between the direct and exchange  contributions in \cref{eq:gamma_m2_exact} in the inset of~\cref{fig:rs_dependence_lowT}, where it is apparent that even at small~$r_s$, the exchange term gives an important nonanalytic correction that quickly saturates to the maximum value~$1/2$, which reflects a fundamental bound~\cite{giuliani_quantum_2005}. At higher~$r_s$, the exchange term thus subtracts half of the direct contribution, such that the direct contribution alone overestimates the correct result by a factor of~$2$. Note that \cref{eq:lowT_even_pred} is exactly half of the direct contribution and thus applies at $r_s \gtrsim 0.5$. Additional results for the asymptotic high-density limit \mbox{$r_s \lesssim (T/T_F) \ll 1$} are listed in Appendix~\ref{app:low_rs_limit}.

Overall, the \mbox{$m=2$} decay rate in \cref{fig:rs_dependence_lowT} increases as \mbox{$\gamma_{m=2} = - 4 \pi T^2 r_s^2 \ln (r_s e^{1+\pi/4}/\sqrt{2})/3 \hbar T_F$} for small interaction strengths \mbox{$(T/T_F)\lesssim r_s\ll 1$}, consistent with the functional dependence at low $r_s$ in Refs.~\cite{alekseev_viscosity_2020,gran_shear_2023-1}, but quickly deviates from this limit and crosses over to a constant \mbox{$\gamma_{m=2} = 2 \pi T^2/3\hbar T_F$} at large \mbox{$r_s$}. This behavior is understood from the head-on scattering kinematics that sets the even-parity decay (cf.~\cref{fig:1}), which at low temperatures involves scattering under any angle of the Fermi surface with a general momentum transfer of order \mbox{${\it O}(k_F) = {\it O}(r_s^{-1})$}. For high densities, \mbox{$r_s \ll 1$}, screening is no longer efficient and the matrix element decreases with momentum, where the logarithmic correction reflects the divergence of the unscreened Coulomb interaction at zero (or $2k_F$) momentum transfer, which favors small-angle scattering~\cite{raines21}. In the low-density limit, \mbox{$r_s \gg 1$}, the Coulomb matrix element is approximately constant with a magnitude set by the the wave vector~$k_{\rm TF}$. In this case, the Thomas-Fermi wave vector cancels with the overall magnitude of the interaction, such that the overscreened interaction is a constant that depends on the density of states and is independent of~$r_s$.

In summary, the different dependence of the lowest odd- and even-parity decay rates on the Coulomb interaction strength~$r_s$  provides a direct signature of the distinct microscopic relaxation processes for these modes. Moreover, it allows us to tune the strength of the odd-even effect not only by varying the temperature but also by changing the interaction strength (for example, by doping). Here, even small changes in $r_s$ have a strong effect: For example, an increase from \mbox{$r_s = 1$} to \mbox{$r_s = 2$} increases the difference between the odd and even decay rates in \cref{fig:rs_dependence_lowT}(a) by $40\%$. For 2D GaAs parameters, this corresponds to decreasing the doping density from $3.3\times10^{11}$cm$^{-2}$ to $0.8\times10^{11}$cm$^{-2}$, which is readily accessible in experiments~\cite{keser_geometric_2021}. 
\section{Fermi-liquid kinetic theory}
\label{sec:background}
In this section, we discuss in detail the structure of the collision integral,~\cref{eq:collision_def}, and the formal basis expansion framework used to diagonalize the collision integral and to derive the collective quasiparticle decay rates presented in the previous section. 

The main object of interest in a Fermi-liquid description is the (single-particle electron) quasiparticle distribution function \mbox{$f(t, \mathbf{r},\mathbf{p})$}, the evolution of which is described semiclassically by the transport equation
\begin{equation}
    \lp \pdv{t} + 
    \mathbf{v}
    \cdot \pdv{\mathbf{r}} 
    +\mathbf{F} \cdot \pdv{\hbar\mathbf{p}}\rp f(t, \mathbf{r},\mathbf{p}) = \hat{\mathcal{J}}[f(t, \mathbf{r}, \mathbf{p})] ,
    \label{eq:Boltzmann_start}
\end{equation}
where \mbox{$\mathbf{v}= \hbar \mathbf{p}/m^*$} is the quasiparticle velocity and $\mathbf{F}$ is an external force. The left-hand side contains the streaming term that describes free phase-space evolution, and the right-hand side is the collision integral stated in~\cref{eq:collision_def}, which describes the rate of change of the distribution \mbox{$f(t, \mathbf{r}, \mathbf{p})$} due to two-body collisions with other quasiparticles. Formally, the collision integral \mbox{$\hat{\mathcal{J}}$} depends on the two-particle distribution, which in turn obeys its own dynamical equation dependent on the three-particle distribution, and so on. Assuming that incoming states are uncorrelated, this hierarchy may be truncated at first order, which gives \cref{eq:collision_def} with a scattering matrix
\begin{align}
&W(\mathbf{p}_1', \mathbf{p}_2' | \mathbf{p}_1 \mathbf{p}_2) = 
\frac{2 \pi}{\hbar}\abs{V}^2 \\
&\times (2 \pi)^2 \delta (\mathbf{p}_1+\mathbf{p}_2-\mathbf{p}_1'-\mathbf{p}_2') \delta ( \varepsilon_{\mathbf{p}_1} + \varepsilon_{\mathbf{p}_2} -  \varepsilon_{\mathbf{p}_1'} - \varepsilon_{\mathbf{p}_2'}) \label{eq:W1212} ,
\end{align}
where the delta functions enforce energy and momentum conservation. The collision integral is then a Fermi golden rule expression for the change in the quasiparticle occupation $f$. For a two-component Fermi gas with spin-independent Coulomb interactions, the matrix element in~\cref{eq:W1212} reads
\begin{align}
    \abs{V}^2 =\abs{\mel{\mathbf{p}_1' \mathbf{p}_2'}{V}{\mathbf{p}_1 \mathbf{p}_2}}^2 &= V^2(\mathbf{p}_1' - \mathbf{p}_1 ) + V^2(\mathbf{p}_1' - \mathbf{p}_2 ) \\[1ex]
    & \quad - V(\mathbf{p}_1' - \mathbf{p}_1 )V(\mathbf{p}_1' - \mathbf{p}_2) ,
    \label{eq::Coulomb_matel_channels}
\end{align}
where $V$ is the interaction potential stated in~\cref{eq:Coulomb_matel}. The first two terms in~\cref{eq::Coulomb_matel_channels} are the  direct contribution and the last term is the exchange contribution~\cite{giuliani_quantum_2005}.
 
The next section, \cref{sec:linearized_collision}, discusses the linearized collision integral and its mathematical properties, focusing in particular on its scalar product structure. The subsequent \cref{sec:basis_expansion} presents the basis expansion used to determine the eigenmodes of the linearized collision integral, which set the relaxation rates presented in the previous section. Finally, \cref{sec:convergence} illustrates the rapid convergence of the basis function expansion, and \cref{sec:zeromodes} discusses the zero modes of the collision integral.
\subsection{Linearized collision integral}
\label{sec:linearized_collision}
As discussed in the Introduction, to describe small deviations from the equilibrium Fermi-Dirac distribution $f_0(\mathbf{p})$, we linearize the collision integral as
\begin{align}
    f(t, \mathbf{p}) = f_0(\mathbf{p}) + \delta f(t, \mathbf{p}) ,
\end{align}
where we further separate a Fermi factor from the small deviation~$\delta f$ and  parametrize it by a deviation function~$\psi$, cf.~\cref{eq:f_pert_def}. To linearize the collision integral, \cref{eq:collision_def}, it is helpful to introduce the center-of-mass and relative momentum variables for the initial-state momenta,
\begin{equation}
    \mathbf{P} = \mathbf{p}_1+ \mathbf{p}_2, \quad \mathbf{q} = \frac{\mathbf{p}_1-\mathbf{p}_2}{2},
    \label{eq:vars_change}
\end{equation}
and likewise for the final-state momenta \mbox{$\mathbf{P}'$ and $\mathbf{q}'$}. The delta functions in the matrix element in \cref{eq:W1212} then imply a conservation of the center-of-mass momentum \mbox{$\mathbf{P} = \mathbf{P}'$} and the magnitude of the relative momentum, \mbox{$q'= q$}. Furthermore, the product of Fermi factors in the collision integral may be written as
\begin{align}
    &F_{121'2'}
    \equiv f_0(\mathbf{p}_1) f_0(\mathbf{p}_2) [1 - f_0(\mathbf{p}_1')] [1-f_0(\mathbf{p}_2')] \\
    &= \frac{1}{4\lsb \cosh (X) + \cosh (\xi \mathbf{P}\cdot \mathbf{q}) \rsb \lsb \cosh (X) + \cosh (\xi \mathbf{P}\cdot \mathbf{q}') \rsb} 
    \label{eq:F1212}
\end{align}
with \mbox{$\xi = \hbar^2 \beta/2 m^*$} and
\begin{equation}
    X = \beta \lp \frac{\hbar^2 P^2}{8 m^*} + \frac{\hbar^2 q^2}{2 m^*} - \mu\rp.
    \label{eq:xi_w_def}
\end{equation}
Using the parametrization \cref{eq:f_pert_def}, this brings the collision integral to the form
\begin{align}
    \hat{\mathcal{J}}[\delta f(\mathbf{p}_1)] =& -\frac{m^*}{4 \pi \hbar^3} \int \frac{\d \mathbf{p}_2}{(2 \pi)^2} \int\d \Omega \abs{V}^2 F_{121'2'}\\
    &\times\lsb \psi(\mathbf{p}_1) + \psi(\mathbf{p}_2) - \psi(\mathbf{p}_1') - \psi(\mathbf{p}_2')\rsb ,
    \label{eq:collision_psi}
\end{align}
where $\Omega$ denotes the angle between the relative momenta $\mathbf{q}$ and $\mathbf{q}'$ before and after the collision. 

Even in linearized form, the collision integral remains a computational challenge. The computation of the electron-electron lifetime is therefore often limited to the relaxation time approximation~\cite{giuliani_quantum_2005,baym_landau_1991,pines_theory_2018}, or to the quasiparticle lifetime obtained from self-energy methods in the zero- or low-temperature limit~\cite{chaplik_energy_1971,hodges_effect_1971,giuliani_lifetime_1982}. For a direct diagonalization of the collision integral, we introduce a scalar product that is induced by the Fermi-Dirac distribution, 
\begin{align}
    \braket{\chi'}{\chi} &= \lambda_T^2 \int \frac{\d^2\mathbf{p}}{(2 \pi)^2} f_0(\mathbf{p}) [1-f_0(\mathbf{p})] \bar{\chi}'(\mathbf{p}) \chi (\mathbf{p}) \\
    &= \int_{-\beta \mu}^\infty \frac{\d w}{4 \cosh^2(w/2)}\bar{\chi}'(w)\chi (w),
    \label{eq:scalar_prod_def}
\end{align}
where the latter equality is valid for isotropic functions and we define the dimensionless energy variable
\begin{align} \label{eq:definitionw}
w = \beta\lp\frac{\hbar^2 p^2}{2 m^*}-\mu\rp .
\end{align}
Analytical results for the scalar product of simple monomials are collected in Appendix~\ref{app:scalarProduct}. Note that the first line in the definition~\eqref{eq:scalar_prod_def} applies equally well for general (anisotropic) dispersions through the Fermi-Dirac functions, while the second line is specific to a parabolic dispersion. In the absence of external forces or thermal gradients, the streaming term in~\cref{eq:Boltzmann_start} vanishes and we obtain the eigenvalue problem
\begin{equation}
    \hat{\mathcal{L}}[\psi(\mathbf{p})] = \frac{- \hat{\mathcal{J}}[\delta f(\mathbf{p})]}{f_0(\mathbf{p}) (1- f_0(\mathbf{p}))} = \gamma \, \psi(\mathbf{p}) 
    \label{eq:L_gamma_def}
\end{equation}
for the linearized collision operator $\hat{\mathcal{L}}$, where an eigenmode $\psi$ decays exponentially with time as \mbox{$\psi(t, \mathbf{p}) = e^{-\gamma t} \psi(\mathbf{p})$}.
With respect to the scalar product Eq.~\eqref{eq:scalar_prod_def}, matrix elements of the operator $\hat{\mathcal{L}}$ then take the form
\begin{align}
    \label{eq:L_matel}
    \mel{\chi}{\hat{\mathcal{L}}}{\psi} &= \frac{m^* \lambda_T^2}{16 \pi \hbar^3} \int \frac{\d \mathbf{P}\, \d \mathbf{q} \, \d \Omega}{(2 \pi)^4} \abs{V}^2 F_{121'2} \\
    &\quad \times \lsb \bar{\chi}(\mathbf{p}_1) + \bar{\chi}(\mathbf{p}_2) - \bar{\chi}(\mathbf{p}_1') - \bar{\chi}(\mathbf{p}_2')\rsb\\
    &\quad \times \lsb \psi(\mathbf{p}_1) + \psi(\mathbf{p}_2) - \psi(\mathbf{p}_1') - \psi(\mathbf{p}_2')\rsb,
\end{align}
where we have implemented the change of variables in~\cref{eq:vars_change} and used the (anti)symmetry under \mbox{$1 \leftrightarrow 2$} \mbox{($1 2 \leftrightarrow 1' 2'$)} to symmetrize the term in the first square bracket, which introduces an additional factor $1/4$. Written in this way, \cref{eq:L_matel} defines a manifestly positive semidefinite Hermitian operator [which follows from \mbox{$F_{121'2'}>0$}, as seen in~\cref{eq:F1212}], which implies that all eigenvalues $\gamma$ are real and non-negative, as required for decay rates. Note that the mathematical structure of the linearized collision integral discussed here including the positive semidefiniteness is not specific to parabolic dispersions but also applies for general anisotropic dispersions.
\subsection{Basis expansion}\label{sec:basis_expansion}
We systematically describe quasiparticle deformations in a given angular sector with angular index~$m$ by expanding the residual radial basis function as
\begin{equation}
    \psi_m (p) = \sum_{i=1}^N c_i u_i(w)
    \label{eq:basis_expansion}
\end{equation}
up to a basis dimension $N$, where the basis polynomials $u_i(w)$ are orthonormal with respect to the scalar product~\cref{eq:scalar_prod_def}; i.e., they obey \mbox{$\langle u_i | u_j \rangle = \delta_{ij}$}. Using the analytical result for the scalar products between monomials listed in Appendix~\ref{app:scalarProduct}, it is straightforward to determine these basis polynomials by a Gram-Schmidt procedure. Note that the presence of the Fermi factors in the scalar product Eq.~\eqref{eq:scalar_prod_def} implies that the basis polynomials depend on temperature. In particular, in the low-temperature limit \mbox{$T/T_F \to 0$}, we obtain
\begin{align}
    u_1(w) &= 1, \\
    u_2(w) &= \frac{\sqrt{3}}{\pi} w, \\
    u_3(w) &= \frac{3\sqrt{5}}{4 \pi^2}w^2-\frac{\sqrt{5}}{4},\\
    u_4(w) &= \frac{5\sqrt{7}}{12 \pi^3}w^3-\frac{7\sqrt{7}}{12 \pi}w, \\
    u_5(w) &= \frac{35}{64 \pi^4}w^4-\frac{65}{32 \pi^2}w^2+ \frac{27}{64},\\
    u_6(w) &= \frac{21\sqrt{11}}{320 \pi^5}w^5-\frac{49\sqrt{11}}{96 \pi^3}w^3 + \frac{407\sqrt{11}}{960\pi}w,
    \label{eq:lowT_exact_polys}
\end{align}
and so on. As discussed in the Introduction, the constant basis function $u_1(w)$ describes a rigid shift in the chemical potential, and higher corrections account for a broadening of the Fermi surface. In the high-temperature limit, the basis polynomials are Laguerre (or Sonine) polynomials \mbox{$u_\ell (w) = e^{-\beta\mu/2} L_{\ell-1}(w+\beta \mu)$}.  

What remains for an evaluation of \cref{eq:L_matel} is then a matter of evaluating the \mbox{$N\times N$} matrix of scalar products between different matrix elements,
\begin{equation}
    \mathcal{M}^{m}_{ij} = \langle u_i e^{i m \theta} \vert \hat{\mathcal{L}} \vert u_j e^{i m \theta} \rangle ,
    \label{eq:matrix_elements}
\end{equation}
where the matrix elements are defined in \cref{eq:L_matel}. Note that the matrix elements themselves do not depend on~$N$. 
The $\{\gamma_{m}^{(N)}\}$ converge to the true decay rates $\{ \gamma_{m} \}$ from above as the basis size $N$ is increased. 
\subsection{Numerical implementation and convergence}
\label{sec:convergence}

\begin{figure}[t]
    \centering
    \includegraphics{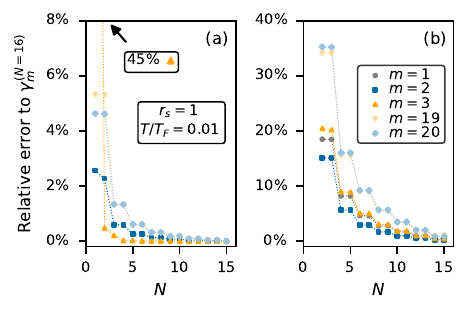}
    \caption{
    Convergence of the numerical results for selected (a) lowest eigenmodes and (b) second-lowest modes.  Shown is the error of the decay rate $\gamma_m^{(N)}$ computed with basis size $N$ relative to $\gamma_m^{(N=16)}$ at low temperature \mbox{$T/T_F=0.01$} and interaction strength \mbox{$r_s = 1$}. 
    }
    \label{fig:convergence1}
\end{figure}

As discussed, we determine the eigenvalues~\mbox{$\{\gamma_{m}^{(N)}\}$} of the linearized collision integral by diagonalizing the \mbox{$N\times N$} coefficient matrix~\cref{eq:matrix_elements}. Here, the numerical challenge is the evaluation of the matrix elements~\cref{eq:L_matel}, which are multidimensional integrals. Furthermore, at low temperatures, the integrand is strongly peaked in a small subdomain of the integration region due to Pauli blocking [enforced by the Fermi factors $F_{121'2'}$, \cref{eq:F1212}]. In our calculations, we use an adaptive multidimensional integration routine contained in the Cuba library~\cite{hahn_cubalibrary_2005}, which was developed to evaluate higher-loop Feynman diagrams in high-energy physics (for applications to diagrammatic calculations in a condensed matter context, see Refs.~\cite{hofmann14,throckmorton15,holder15}). The computation speed at low temperatures is greatly improved by using the \textit{Divonne} algorithm, which allows the Monte Carlo sampling to bias the integration points in regions where the integrand is peaked. This allows us to achieve rapid convergence on standard computers. We provide further details of the numerical implementation and a dimensionless form of~\cref{eq:L_matel} in Appendix~\ref{app:numerical_details}. In this section, we illustrate the convergence of the eigenvalues with increasing basis dimension $N$.

\Cref{fig:convergence1} illustrates the convergence for selected eigenmodes $\gamma_m^{(N)}$ as a function of basis dimension~$N$, where \cref{fig:convergence1}(a) shows the lowest eigenvalue  for \mbox{$m=2,3,19,$ and $20$}, and \cref{fig:convergence1}(b) shows the second-lowest eigenvalue for \mbox{$m=1,2,3,19,$ and $20$}. [The lowest eigenvalue for \mbox{$m=1$} is excluded from \cref{fig:convergence1}(a) since it is a zero mode. It is discussed separately in the next section, \cref{sec:zeromodes}.] We plot the relative deviation of $\gamma_m^{(N)}$ compared to $\gamma_m^{(N=16)}$, which we take as the converged value, as a function of~$N$. Data shown are for \mbox{$r_s=1$} and \mbox{$T/T_F = 0.01$}, where we do not see a pronounced difference in the convergence rate for other temperatures below the Fermi temperature. Lower (higher) interaction strengths lead to a minor improvement (reduction) of the convergence. As is apparent from the figure, for all eigenvalues we see uniform rapid convergence, which is fastest for the lowest eigenvalue. In particular, for \mbox{$N=6$}, the decay rates of the $m=2$ mode are within $1\%$ and $6 \%$ of the converged value in the lowest and second-lowest sector, respectively, and we use this basis dimension to generate the results shown in~\cref{fig:allrates_temperature,fig:mscaling_lowT}. For the remaining figures, we have used \mbox{$N=10$}. The rapid convergence of these results demonstrates that our choice of orthogonal basis polynomials in \cref{sec:basis_expansion} is very well suited to diagonalize the Fermi-liquid collision integral. An open question left for future work is if a different choice of basis functions could improve the convergence even further. 

An interesting feature of \cref{fig:convergence1}(a) is that the onset of the $T^4$-scaling regime and the odd-even effect is mirrored in the convergence pattern: For the \mbox{$m=3$} mode, which admits a scaling exponent \mbox{$\alpha \approx 4$} for \mbox{$T=0.01T_F$} [cf.~the inset of \cref{fig:allrates_temperature}(a)], the first basis function alone (\mbox{$N=1$}) provides a poor description that differs by \mbox{$45\%$} [as marked in the inset of \cref{fig:convergence1}(a)], while including the first two basis functions (\mbox{$N=2$}) gives a dramatic improvement in accuracy. This is in line with the explicit calculation of the lowest matrix element at low temperatures in Appendix~\ref{app:lowTdetails}, which vanishes for odd $m$. By contrast, the \mbox{$m=19$} mode, which still follows Fermi-liquid scaling at \mbox{$T=0.01 T_F$}, exhibits a similar convergence pattern to the even modes with good convergence even for \mbox{$N=1$}. 
\subsection{Zero modes}
\label{sec:zeromodes}
The linearized collision integral~\cref{eq:collision_psi} possesses four zero modes, which are associated with the conservation of particle number, energy, and the $x$ and $y$ components of the total momentum in binary collisions:
\begin{align}
    \mathcal{N} &= \int \frac{\d \mathbf{p}}{(2 \pi)^2} f(t,\mathbf{p}) , \\
    E &= \int \frac{\d \mathbf{p}}{(2 \pi)^2} f(t,\mathbf{p}) \varepsilon (\mathbf{p}) , \\
    \mathbfcal{P} &= \int \frac{\d \mathbf{p}}{(2 \pi)^2} f(t,\mathbf{p}) \hbar \mathbf{p} . \\
\end{align}
Using the kinetic equation~\cref{eq:Boltzmann_start} and symmetrizing, the time derivative of these quantities is expressed in terms of the linearized collision integral,
\begin{align}
    \frac{d}{dt} \mathcal{O} = \langle A({\bf p}) | \mathcal{\hat{L}}[\psi] \rangle
\end{align}
for a general deformation $\psi$, where \mbox{$\mathcal{O} = \{\mathcal{N}, E, \mathbfcal{P}\}$} and \mbox{$A(\mathbf{p}) = \{1, \, \varepsilon(\mathbf{p}), \, \hbar\mathbf{p}\}$}, respectively. As can be seen directly from the definition~\cref{eq:L_matel}, this scalar product vanishes due to energy and momentum conservation in binary collisions~\cite{tong_kinetic_2012}. From the hermicity of the linearized collision integral, it follows then that perturbations of the form \mbox{$\psi(\mathbf{p}) = A(\mathbf{p})$} are zero modes of the linearized collision integral. As already discussed in \cref{sec:temperaturedependence}, the two zero modes corresponding to particle and energy conservation, which take the form \mbox{$\psi(\mathbf{p}) = 1$} and \mbox{$\psi(\mathbf{p}) = \beta \varepsilon(\mathbf{p}) = w + \beta\mu$}, respectively, have \mbox{$m=0$} symmetry. These modes include the first two basis functions $u_1$ (which is constant in $w$) and $u_2$ (which is linear in $w$) at any temperature, which implies that the first two rows and columns of the coefficient matrix $\mathcal{M}_{ij}^{m=0}$ are identically zero.

\begin{figure}[t]
    \centering
    \includegraphics{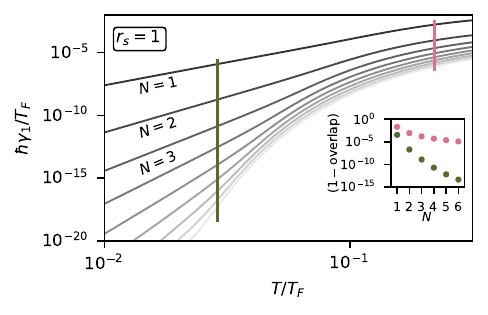}
    \caption{Relaxation rate of the lowest current mode $\gamma_{m=1}^{(N)}$ for finite basis dimension~\mbox{$N=1,\ldots,9$} as a function of temperature $T/T_F$. Lines from top to bottom (darker to lighter colors) show the decay rate with increasing basis dimension~$N$. The mode rapidly converges to the current zero mode with increasing~$N$. The inset illustrates the overlap between the eigenvector of $\mathcal{M}^{m=1}_{ij}$ for a given basis size~$N$ and the first~$N$ elements of the true zero mode $\hbar\mathbf{p}$ (both modes are normalized to unity). At both temperatures shown, \mbox{$T/T_F=0.03$} (dark green points) and \mbox{$T/T_F = 0.22$} (pink points), we find rapid convergence.}
    \label{fig:m1}
\end{figure}

The situation is slightly more involved for the two current zero modes, which take the form \mbox{$\psi({\bf p}) = p e^{\pm i\theta}$} in the \mbox{$m=\pm 1$} sector, and which have a nonanalytic dependence on the dimensionless energy variable $w$ since \mbox{$p \sim \sqrt{w + \beta \mu}$}. Given our choice of basis functions, which are polynomials in the energy deviation $w$, there is thus an overlap of the zero mode with basis polynomials of any order. For this reason, the diagonalization of the coefficient matrix~\cref{eq:matrix_elements} at any finite $N$ will not give an eigenvalue that is identically zero but rather a finite lowest eigenvalue that decreases without lower bound as $N$ is increased. We illustrate this in \cref{fig:m1}, which shows the lowest eigenvalue $\gamma_{m=1}^{(N)}$ for \mbox{$r_s=1$} as a function of temperature, where lines from top to bottom correspond to an increasing basis dimension $N$. The decay is most pronounced at low temperatures, again indicating the suitable choice of basis functions. In addition, we show in the inset the overlap of the numerically obtained eigenvector with the true zero mode as a function of $N$, which is very large even at small $N$.
\section{Conclusion and Outlook}\label{sec:conclusion}
In summary, we have provided a comprehensive discussion of nonequilibrium collective quasiparticle relaxation in two-dimensional electron gases with a screened Coulomb interaction, which is made possible by the exact diagonalization of the linearized Fermi-liquid collision integral. The focus of our work is the description of anomalously long-lived odd-parity modes with decay rates that are much more strongly suppressed than predicted by Fermi-liquid theory. Just as for regular Fermi-liquid decay, the lifetimes of these modes are increased by Pauli blocking, but their microscopic relaxation mechanism differs: Instead of variable-angle head-on collisions, relaxation proceeds by repeated small-angle scattering on the Fermi surface, which is manifest by a strong dependence \mbox{${\it O}(m^4)$} on the angular harmonic index $m$. We find that the parity-effect in the quasiparticle decay rates is restricted to the longest-lived eigenmode in a given odd-parity channel, and higher modes show standard Fermi-liquid behavior. There is a finite set of order \mbox{${\it O}(\sqrt{T_F/T})$} of isolated long-lived eigenmodes, where each mode has a separate crossover temperature below which it decouples from the Fermi-liquid continuum. We show furthermore that the crossover point also depends on the Coulomb interaction strength $r_s$, and that the separation between odd- and even-parity decay rates is strongly dependent on the interaction strength, which reflects the different relaxation processes. 
Our numerical calculations demonstrate that the odd-even anomaly is not just an asymptotic result valid at unrealistically small temperatures, but should be readily accessible in current experiments.

While our work quantifies the odd-even effect and systematically determines the collisional eigenmodes as well as their decay rates, a natural follow-up question is how the long-lived modes affect the dynamics of the electron gas, in particular how they can be incorporated in a hydrodynamic description. In view of the strong separation of an isolated set of odd-parity modes reported here, such a description should be robust, and our results provide input parameters for an effective description in terms of hydrodynamic densities and odd-parity modes. Here, recent work has provided a systematic description of tomographic transport in finite geometries using a Knudsen expansion~\cite{benshachar25a,benshachar25b}. This expansion reveals significant deviations for tomographic flow compared to a hydrodynamic description in the form of rarefaction corrections to the bulk Stokes-Ohm equation, an anomalously large extended tomographic boundary layer, and additional velocity slip at the boundaries. These effects must be included when describing electron flow in realistic devices. Moreover, even weak magnetic fields will interfere with the anomalously long-lived odd modes and suppress tomographic transport effects, thus providing a hallmark signature of this regime~\cite{rostami25}. A further tantalizing prospect given the strong suppression and near conservation of odd-parity modes is that methods from integrable systems that incorporate an infinite number of conserved modes in a generalized hydrodynamic framework could be applied to the present problem~\cite{essler23}. 

On a technical level, we have presented a numerically exact method to diagonalize the Fermi-liquid collision integral, which is necessary since other means of computing the electron relaxation, for example, using self-energy methods, are not sensitive to the odd-even effect. Such a direct solution is often avoided due to the numerical complexity, and our work provides an efficient algorithm to address this problem. In particular, the calculation of general linear-response transport coefficients, which as discussed in the Introduction are linked to particular superpositions of the eigenmodes discussed here, is an immediate prospect for future work. With regard to the odd-even effect, with the disparity between odd and even modes most pronounced for \mbox{$m \lesssim \sqrt{T_F/T}$}, the focus should be on transport coefficients that couple strongly to harmonics with low values of $m$ that are not dominated by an even Fermi-liquid mode. Likewise, the generalization of the present description to anisotropic Fermi surfaces is a prospect for future work.

\begin{acknowledgments}
This work is supported by Vetenskapsr\aa det (Grants No.~2020-04239 and No.~2024-04485), the Olle Engkvist Foundation (Grant No.~233-0339), the Knut and Alice Wallenberg Foundation (Grant No.~KAW 2024.0129), Nordita (J.H.) and the Chalmers' Nano Area of Advance under its Excellence Ph.D.\@ program (E.N.). The computations were enabled by resources provided by the National Academic Infrastructure for Supercomputing in Sweden (NAISS) at Linköping University partially funded by the Swedish Research Council through Grant Agreement No.~2022-06725.
\end{acknowledgments}
\appendix
\section{LOW-TEMPERATURE APPROXIMATION OF EVEN-PARITY RELAXATION RATES}
\label{app:lowTdetails}

\begin{figure}[b]
    \centering
    \includegraphics{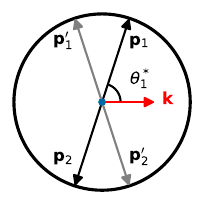}
    \caption{Example of a head-on collision process (\mbox{$\mathbf{P} = 0$}) at low temperatures, where scattering momenta are at the Fermi surface. At zero energy transfer $\omega = 0$, $\cos \theta_1^* = k/2p_1$.}
    \label{fig:head-on}
\end{figure}

In this appendix, we compute the asymptotic low-temperature scaling of the lowest even-parity decay rates analytically. To this end, we evaluate the matrix elements of the linearized collision integral, 
\begin{equation}
    \gamma_{m \text{ even}} \approx  \mel{\psi_\star}{\hat{\mathcal{L}}}{\psi_\star}
\end{equation}
for the leading perturbation \mbox{$\psi_\star(\mathbf{p}) = e^{im \theta}$}, i.e., the lowest-order basis function that describes a rigid shift of the chemical potential. As discussed in \cref{sec:convergence} and highlighted in \cref{fig:convergence1}, this provides a good approximation of the decay rate for modes with even $m$. 
Explicitly, the collision matrix element reads
\begin{align}
    &\mel{\psi_\star}{\hat{\mathcal{L}}}{\psi_\star} = \frac{2 \pi \lambda_T^2}{4 \hbar} \iiiint \frac{\d \mathbf{p}_1 \d \mathbf{p}_2 \d \mathbf{p}_1' \d \mathbf{p}_2'}{(2\pi)^8} \abs{V}^2 \\
    &\times \delta \lp \varepsilon_{\mathbf{p}_1} + \varepsilon_{\mathbf{p}_2} -  \varepsilon_{\mathbf{p}_1'} - \varepsilon_{\mathbf{p}_2'}\rp (2 \pi)^2 \delta (\mathbf{p}_1+\mathbf{p}_2-\mathbf{p}_1'-\mathbf{p}_2') \\
    &\times F_{121'2'} \abs{\psi_\star(\mathbf{p}_1) + \psi_\star(\mathbf{p}_2) - \psi_\star(\mathbf{p}_1') - \psi_\star(\mathbf{p}_2')}^2 .
    \label{eq:appendixcollision}
\end{align}
Instead of the center-of-mass and relative momentum variables used in the main text, here we choose as integration variables the momentum and energy transfer, splitting the energy and momentum delta functions as~\cite{raines21}
\begin{align}
    &\delta \lp \varepsilon_{\mathbf{p}_1} + \varepsilon_{\mathbf{p}_2} -  \varepsilon_{\mathbf{p}_1'} - \varepsilon_{\mathbf{p}_2'}\rp \\
    &\quad = \int \d \omega\,  \delta \lp \varepsilon_{\mathbf{p}_1} -  \varepsilon_{\mathbf{p}_1'} - \omega\rp \delta \lp  \varepsilon_{\mathbf{p}_2} - \varepsilon_{\mathbf{p}_2'} + \omega \rp,\\ 
    &\delta (\mathbf{p}_1+\mathbf{p}_2-\mathbf{p}_1'-\mathbf{p}_2') \\
    &\quad = \int \d \mathbf{k}\, \delta (\mathbf{p}_1-\mathbf{p}_1' - \mathbf{k}) \delta(\mathbf{p}_2 - \mathbf{p}_2' + \mathbf{k}),
\end{align}
where \mbox{$\mathbf{k} = {\bf p}_1 - {\bf p}_1' = {\bf p}_2' - {\bf p}_2$} is the momentum transfer and \mbox{$\abs{\omega} \sim T \ll T_F$} is the energy transfer. The delta functions then constrain the final momenta \mbox{$\mathbf{p}_1' = \mathbf{p}_1-\mathbf{k}$} and \mbox{$\mathbf{p}_2' = \mathbf{k}+\mathbf{p}_2$}. In the following, we express the angles $\theta_1$ and $\theta_2$ of the remaining initial momenta ${\bf p}_1$ and ${\bf p}_2$ with reference to the direction of the momentum transfer $\mathbf{k}$, as illustrated in \cref{fig:head-on}.

The integral over the initial angles $\theta_1$ and $\theta_2$ is evaluated using the energy delta functions,
\begin{align}
    \int_0^{2\pi} \d \theta_1 \, \delta \lp \varepsilon_{\mathbf{p}_1} -  \varepsilon_{\mathbf{p}_1'} - \omega\rp [\dots] &= \sum_{\theta_1^*} \frac{1}{v_1 \hbar k \abs{\sin(\theta_1^*)}} [\dots], \\
    \int_0^{2\pi} \d \theta_2 \, \delta \lp \varepsilon_{\mathbf{p}_2} -  \varepsilon_{\mathbf{p}_2'} + \omega\rp [\dots] 
    &= \sum_{\theta_2^*} \frac{1}{v_2 \hbar k \abs{\sin(\theta_2^*)}} [\dots] ,
    \label{eq:deltaSums}
\end{align}
where we used \mbox{${\mathbf{p}_i'}^2 = (\mathbf{k}-\mathbf{p}_i)^2$} and \mbox{$v_i = m^* \hbar p_i$}. Each sum runs over the two angles $\theta^\star_i$ for which the argument in the energy delta functions vanishes, i.e., which fulfill
\begin{align}
    \cos(\theta_1^*) &= \frac{k}{2 p_1} + \frac{\omega}{v_1 k}, \\
    \cos(\theta_{2}^*) &= - \frac{k}{2 p_2} + \frac{\omega}{v_2 k}.
    \label{eq:cosThetaStar}
\end{align}
Energy and momentum conservation in combination with the Fermi surface constraint give two different scattering combinations. The first one is a direct or exchange scattering process with \mbox{$k=0$} or \mbox{$k = 2k_F$}, respectively, where \mbox{$\mathbf{p}_1 = \mathbf{p}_1', \mathbf{p}_2 = \mathbf{p}_2'$} (or interchanged final momenta). For these configurations, the matrix element~\cref{eq:appendixcollision} of the collision integral vanishes, as expected and already discussed in the Introduction. The second possible process is head-on scattering with \mbox{$\mathbf{p}_1 = -\mathbf{p}_2, \mathbf{p}_1' = -\mathbf{p}_2'$} (\cref{fig:head-on}). In other words, only two of the four terms that come from combining the sums in~\cref{eq:deltaSums} actually contribute, which gives a factor of $2$.

For $\mathbf{k}$ along the $x$ axis, if $\mathbf{p}_1$ is in the upper (right-hand) plane, $\mathbf{p}_1'$ must be in the upper (left-hand) plane (cf.~\cref{fig:head-on}). Furthermore, the relation \mbox{$ \mathbf{k} = \mathbf{p}_1 - \mathbf{p}_1' = \mathbf{p}_1 + \mathbf{p}_2'$} implies that \mbox{$\theta_{2'} = - \theta_1$}, so \mbox{$\theta_{1'} = \theta_{2'} + \pi = \pi - \theta_1$}. The factor in the matrix element of \cref{eq:appendixcollision} becomes (for even $m$)
\begin{align}
    &\abs{\psi_\star(\mathbf{p}_1) + \psi_\star(\mathbf{p}_2) - \psi_\star(\mathbf{p}_1') - \psi_\star(\mathbf{p}_2')}^2
    \\
    &= \abs{e^{i m \theta_1}\lsb 1 + e^{i m \pi} - e^{i m \pi}e^{-2i m \theta_1} - e^{-2i m \theta_1}\rsb }^2 \\
    &= 16 \sin^2 m \theta_1 ,
\end{align}
but vanishes for odd $m$, as discussed in the Introduction.

\begin{figure}[t]
    \centering
    \includegraphics{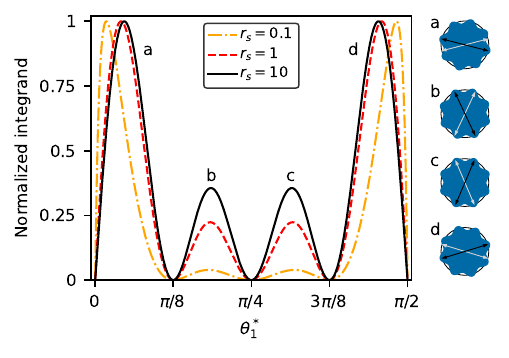}
    \caption{The integrand (normalized by its maximum value) in \cref{eq:lowT_integral} for \mbox{$m = 8$}. As long as $r_s$ is relatively large, the matrix element $\abs{\bar{V}}^2$ is approximately constant over the integration range. The peaks of the integrand roughly correspond to scattering from (black arrow) a peak in the Fermi surface perturbation to (gray arrow) a valley, shown to the right. As~$r_s$ decreases, the contributions from small-angle scattering become increasingly dominant.}
    \label{fig:integrand}
\end{figure}

At this point, the expression for the matrix element~\cref{eq:appendixcollision} reads for even $m$:
\begin{align}
    &\mel{\psi_\star}{\hat{\mathcal{L}}}{\psi_\star} = \frac{\pi \lambda_T^2}{\hbar (2\pi)^6} \lp \int \d\omega \d p_1 \d p_2 \, p_1 p_2 F_{121'2'}\rp \\
    &\quad \times \lp \int_0^\infty\d k \int_0^{2\pi} \d \theta_k \abs{V}^2 \frac{16 \sin^2 m \theta_1^*}{\hbar^2 k v_1 v_2 \abs{\sin \theta_1^* \sin \theta_2^*}} \rp ,
    \label{eq:midway_appendix}
\end{align}
where $F_{121'2'}$ is defined in \cref{eq:F1212} and we neglect the dependence on the momentum transfer $k$ in the product of Fermi functions as well as the dependence of $\theta_1^*$ on the energy transfer as \mbox{$\omega \sim O(T)$}. We transform the wave vectors $p_i$ into dimensionless energies $w_i$ [cf.~\cref{eq:definitionw}] with \mbox{$\hbar^2 p_i \d p_i = m^* \d \varepsilon_i = m^* T \d w_i$}, which in the zero-temperature limit have integration range \mbox{$w \in (-\infty, \infty)$}, and use the exact result for the first factor in \cref{eq:midway_appendix},
\begin{align}
    \int_{-\infty}^\infty \d W \d w_1 \d w_2 F_{121'2'} = \frac{2 \pi^2}{3},
\end{align}
where \mbox{$W =\omega/T$}. For the second factor in \cref{eq:midway_appendix}, we parametrize the $k$ integral using $\theta_1^*$, where \cref{eq:cosThetaStar} gives \mbox{$k = 2 k_F \cos \theta_1^*$} such that \mbox{$0 \leq k \leq 2 k_F$}. Approximating the velocities in the denominator in \cref{eq:midway_appendix} by the Fermi velocities and performing the integral over $\theta_k^*$ gives
\begin{equation}
    \gamma_{m \text{ even}} \approx \frac{4 \pi}{3} \frac{T^2}{\hbar T_F} I_m ,
    \label{eq:gamma_m_low_T}
\end{equation}
where we define the integral
\begin{align}
    I_m &=  \int_0^{2 k_F} \d k \,\frac{\sin^2 m \theta_1^*}{k \sin^2  \theta_1^*} \abs{\bar{V}}^2  \\
    &= \int_0^{\pi/2} \d \theta_1^* \frac{1 - \cos 2 m \theta_1^*}{\sin 2 \theta_1^*} \abs{\bar{V}}^2
    \label{eq:lowT_integral}
\end{align}
with the dimensionless interaction potential \mbox{$\bar{V} = V/\lambda_T^2 T$}, and we  use \mbox{$\abs{\sin \theta_2^*} = \abs{\sin \theta_1^*}$}.

For a constant matrix element $\bar{V}$, the integrand peaks roughly at angles $\theta_1^*$ where the ingoing (\mbox{$\mathbf{p}_1 = -\mathbf{p}_2$})  and outgoing (\mbox{$\mathbf{p}_1' = - \mathbf{p}_2'$}) momenta connect a peak and a valley in the Fermi surface deformation. The fact that the maxima are not exactly located at \mbox{$\theta_1^\star = \tfrac{(\pi+ 2 \ell)}{2m}$}, \mbox{$\ell = 0, 1, \dots$}, is due to the \mbox{$\sin 2 \theta_1^*$} factor in the denominator from energy conservation. We illustrate the integrand for \mbox{$m=8$} in \cref{fig:integrand}. The Coulomb matrix element changes this picture by enhancing small-angle scattering and suppressing large-angle processes. To see this, we express the matrix element as
\begin{align}
    \abs{\bar{V}}^2 &= \bar{V}(\cos \theta_1^*)^2 + \bar{V}(\sin \theta_1^*)^2 - \bar{V}(\cos \theta_1^*)V(\sin\theta_1^*),
    \label{eq:matel_V_chi}
\end{align}
with the dimensionless form of \cref{eq:Coulomb_matel},
\begin{align}
    \Bar{V}(\Delta \phi) &= 
    \frac{r_s}{2(r_s + \sqrt{2} \Delta\phi)} .
\end{align}
For small $r_s$, the matrix element is sharply peaked near \mbox{$\theta_1^* = 0, \pi/2$} and enhances small-angle scattering. Only at large \mbox{$r_s \gtrsim 10$} is the matrix element roughly constant with \mbox{$\abs{\Bar{V}}^2 = 1/4 + O(1/r_s^{2})$}, and we recover the scenario discussed above. 

\Cref{eq:gamma_m2_exact} in the main text stems from the integral $I_{m=2}$, which can be performed analytically for all $r_s$, as shown in~\cref{eq:direct_exchange_m2}. Closed expressions exist also for \mbox{$m = 4,6,8$}, with decreasing convergence radii in $r_s$, and are not stated here. The parameter $\phi$ in Eq.~\eqref{eq:lowT_even_pred} is obtained from a fit to $I_m$ for the first $200$ values of $m$.

\section{EVEN-PARITY RELAXATION RATES IN THE HIGH-DENSITY LIMIT}
\label{app:low_rs_limit}
\begin{figure}[t]
    \centering
    \includegraphics{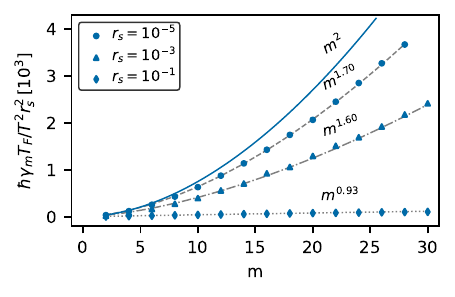}
    \caption{
    Relaxation rates as a function of $m$ at \mbox{$T/T_F = 10^{-4}$} for three different values of the interaction strength $r_s$. The solid blue line corresponds to the decay rate obtained from~\cref{eq:m2_scaling_low_rs}, illustrating that the even modes asymptote to $m^2$ scaling in the \mbox{$r_s \to 0$} limit. Note that the decay rates have been normalized by $r_s^2$ for visibility.
    }
    \label{fig:mscaling_lowT}
\end{figure}
In this appendix, we briefly discuss the asymptotic high-density limit, \mbox{$r_s \to 0$}, for which we can derive a separate analytical result for the even-parity decay rates. 
By expanding the integrand in $I_m$ [cf.~\cref{eq:lowT_integral}] to lowest order in $r_s$ and assuming that the angular integral is cut off by temperature, we regain a logarithmic cutoff in $T/T_F$, which gives
\begin{align}
    I_m \approx & \, \frac{r_{s}^{2}}{4}\lsb m^{2}\lp \frac{3}{2} - \log \frac{T}{T_F}\rp  +(1-m^{2})\lp\gamma_{E} + \log 2 m\rp\rsb\\ 
    &+ \frac{r_s^2 m^2}{4} \frac{T}{T_F} + O\lp \frac{T^2}{T_F^2}, r_s^3\rp ,
    \label{eq:m2_scaling_low_rs}
\end{align}
which we find to be valid for \mbox{$r_s \lesssim T/T_F \ll 1$}. The logarithmic enhancement in $r_s$ seen in~\cref{eq:direct_exchange_m2} therefore drops out at \mbox{$r_s \lesssim (T/T_F)$}, where the decay rate scales only as $r_s^2$.
Hence, the first term (in square brackets) is the direct contribution, and the second ($T$-linear) term is the exchange contribution. The quadratic $m$ dependence is indicative of diffusive dynamics, as can be seen by writing \mbox{$f_{\rm even} = \sum_{m \, {\rm even}} f_m e^{im\theta}$}, which obeys a diffusion equation
\begin{align}
    \partial_t f_{\rm even} &= \gamma_{\rm even} \, \partial_\theta^2 f_{\rm even} ,
\end{align}
with $\gamma_{\rm even}$ set by the quadratic term in \cref{eq:m2_scaling_low_rs}. Such a quadratic-in-$m$ scaling in~\cref{eq:m2_scaling_low_rs} aligns with that in Ref.~\cite{alekseev_viscosity_2020}, but it is attained only for asymptotically small $r_s$ as illustrated  in~\cref{fig:mscaling_lowT}, which shows the numerical relaxation rate at \mbox{$T/T_F = 10^{-4}$} for three small interaction strengths \mbox{$r_s=0.1,0.01,$ and $10^{-5}$}. The scaling of the even modes  agrees very well with the analytical result~\cref{eq:m2_scaling_low_rs}, but the effective $m$ scaling is superdiffusive with local exponents for $m^\beta$ with \mbox{$\beta = 1.7$}, \mbox{$\beta = 1.6$}, and \mbox{$\beta = 0.93$}, respectively, and interpolates between a quadratic and a logarithmic $m$ scaling as $r_s$ is increased. For \mbox{$T/T_F = 10^{-4}$}, the full logarithmic $m$ scaling is reached at \mbox{$r_s \gtrsim 1$}, and much earlier at higher temperatures. We note that at small $r_s$, there is an intermediate regime with linear-in-$m$ scaling of the relaxation rates. Such a scaling is predicted for near-intrinsic graphene, where the relaxation dynamics is interpreted as Lévy flights on the Dirac cone~\cite{kiselev19,kiselev20}.

\section{EXACT RESULTS FOR THE SCALAR PRODUCT}
\label{app:scalarProduct}

In this appendix, we list exact results for the scalar product~\cref{eq:scalar_prod_def} evaluated for monomials of the form~$w^n$, from which all basis functions are composed, where $n$ is an integer and \mbox{$w = \beta(p^2/2m^*-\mu)$}. These formulas both speed up numerical computations and improve the accuracy of our results. In particular, they allow the computation of arbitrary products of polynomials and thus enable the efficient determination of the basis functions. A numerical evaluation is then required only for scalar products that contain nonanalytic expressions (like the velocity \mbox{$\sqrt{w+\beta\mu}$}). Defining \mbox{$\langle w^n \rangle = \langle w^n, 1\rangle$}, we have
\begin{align}
    \langle  w^n \rangle &= 
    \begin{cases} \displaystyle
      - \sum^n_{k=0} \frac{n!}{k!}(\beta\mu)^k {\rm Li}_{n-k}(-e^{-\beta\mu}) & \text{$n$ odd,} \\[4ex]
    \displaystyle - \sum^n_{k=0} \frac{n!}{k!}(-\beta\mu)^k {\rm Li}_{n-k}(-e^{\beta\mu}) & \text{$n$ even,}
    \end{cases}
\end{align}
where ${\rm Li}_n(z)$ is the polylogarithm function. In the zero-temperature limit, this expression simplifies to
\begin{align}
    \langle  w^n \rangle &=  
    \begin{cases} \displaystyle
    0 & \text{$n$ odd,} \\[1ex]
     \displaystyle - 2 n! {\rm Li}_n(-1) & \text{$n$ even.}
    \end{cases}
\end{align}
\section{DIMENSIONLESS FORM OF THE COLLISION INTEGRAL}
\label{app:numerical_details}
In this appendix, we list the dimensionless form of the matrix elements of the collision integral~\cref{eq:L_matel} that is used in our numerical implementation. We express the collision integral, which has dimension of frequency, in units of \mbox{$\hbar/T_F$} and choose dimensionless wave vectors,
\begin{equation}
    \Bar{P} = \frac{\lambda_T}{\sqrt{4 \pi}}P = \sqrt{\frac{\hbar ^2 \beta}{2 m^*}}P , 
    \label{eq:dimless_momentum_def}
\end{equation}
as well as a dimensionless interaction potential \mbox{$V(k) = \lambda_T^2 T \bar{V}$}. In addition, we transform to polar coordinates for the center-of-mass and relative momentum. The center-of-mass angle integral is trivial as neither the Fermi factors nor the Coulomb matrix element depend on~$\theta_P$. We remark that this integral ensures that the matrix elements are diagonal in the angular harmonic component $m$, i.e., \mbox{$\mathcal{M}^{mm'} \sim \delta^{mm'}$}. Note that this diagonal structure in the angular component is no longer present for anisotropic Fermi surfaces. Here, matrix elements of the collision integral with a general $C_p$-symmetric Fermi surface will couple angular modes modulo $p$, i.e, $m$ and $m\pm p$, $m\pm 2p$, and so on. Aligning the center-of-mass momentum with the coordinate axis and denoting the angles between ${\bf P}$ and the relative momenta ${\bf q}$ and ${\bf q}'$ by~$\theta_q$ and $\theta_{q'}$, respectively, we introduce the notation
\begin{align}
    &\quad \quad  \abs{\langle\Bar{V}(\Bar{q}, \Bar{q}')\rangle}^2 = \mathcal{V}_1^2 + \mathcal{V}_2^2 - \mathcal{V}_1 \mathcal{V}_2 , \\
    \mathcal{V}_1 & = \bar{V}\lp \Bar{q} \sqrt{2 \abs{1 - \cos \theta_q \cos \theta_{q'} - \sin \theta_q \sin \theta_{q'}}}\rp, \\ 
    \mathcal{V}_2 &= \bar{V}\lp \Bar{q} \sqrt{2 \abs{1 + \cos \theta_q \cos \theta_{q'} + \sin \theta_q \sin \theta_{q'}}}\rp ,
\end{align}
to obtain the following dimensionless form of the collision integral:
\begin{widetext}
\begin{align}
    \mathcal{M}_{ij}^m &= \lp \frac{T}{T_F}\rp\frac{1}{4 \pi} \int_0^\infty \Bar{P} \d \Bar{P} \int_0^\infty \Bar{q} \d \Bar{q} \int_0^{2 \pi} \d \theta_{q} \int_0^{2 \pi} \d \theta_{q'} \abs{\langle\Bar{V}(\Bar{q}, \Bar{q}')\rangle}^2\\
    \times & \frac{1/2}{\cosh (w) + \cosh (\Bar{P} \Bar{q} \cos \theta_q)} \frac{1/2}{\cosh (w) + \cosh (\Bar{P} \Bar{q}' \cos \theta_{q'})} \\
    \times  &\lsb \Bar{u}_j(w + \Bar{P}\Bar{q}\cos \theta_q) e^{- i m \theta_1} + \Bar{u}_j(w - \Bar{P}\Bar{q}\cos \theta_q) e^{- i m \theta_2} - \Bar{u}_j(w + \Bar{P}\Bar{q}'\cos \theta_{q'}) e^{- i m \theta_1'} - \Bar{u}_j(w + \Bar{P}\Bar{q}'\cos \theta_{q'}) e^{- i m \theta_2'}\rsb\\
    \times &\lsb u_i(w + \Bar{P}\Bar{q}\cos \theta_q) e^{i m \theta_1}  + u_i (w - \Bar{P}\Bar{q}\cos \theta_q) e^{i m \theta_2} - u_i (w + \Bar{P}\Bar{q}'\cos \theta_{q'}) e^{i m \theta_1'} - u_i (w - \Bar{P}\Bar{q}'\cos \theta_{q'}) e^{ i m \theta_2'}\rsb, 
    \label{eq:dimless_Lmatel}
\end{align}
\end{widetext}
where \mbox{$w = \Bar{P}^2/4 + \Bar{q}^2 - \beta\mu$} [cf.~\cref{eq:xi_w_def}]. The basis functions have here been written in their explicit form, 
\begin{equation}
    \psi(\mathbf{p}) = u^{(N)}_i (u) e^{i m \theta},
\end{equation}
where $u^{(N)}_i$ is the $i$th basis polynomial (\mbox{$i \leq N$}), cf. \cref{sec:basis_expansion}. Finally, the angles \mbox{$\theta_1, \theta_2$} are related to the integration variables as
\begin{align}
    \theta_{1,2} &= \theta_P +f(\Bar{P}/2, \pm \Bar{q}, \theta_{q}),\\
    \theta'_{1,2} &= \theta_P + f(\Bar{P}/2, \pm \Bar{q}, \theta_{q'}),\\
    f(\Bar{P}, \Bar{q},\theta) &= \arctan\lp\frac{\Bar{q}\sin \theta}{\Bar{P}+\Bar{q}\cos\theta}\rp .
\end{align}
As discussed in \cref{sec:background}, $\hat{\mathcal{L}}$ is a Hermitian operator, so \mbox{$N(N+1)/2$} different overlap integrals need to be computed in order to obtain the full \mbox{$N\times N$} matrix~$\mathcal{M}^m_{ij}$. 

Note that at low temperatures, the product of Fermi functions is a sharply peaked function in the radial coordinate \mbox{$\rho^2 = \Bar{P}^2/4 + \Bar{q}^2$}. For this reason, it is helpful in numerical implementations to perform a final change of integral variables as \mbox{$\Bar{P} = 2\rho\cos\theta$} and \mbox{$\Bar{q} = \rho\sin\theta$}. The advantage of this parametrization is that numerical adaptive Monte Carlo methods (for example, the \textit{Divonne} algorithm; see Ref.~\cite{hahn_cubalibrary_2005}) may be supplemented with a peak-finder routine, allowing greater numerical accuracy at low temperatures.
\bibliography{bib_kinetic}

\end{document}